\documentclass[epj]{svjour}
\usepackage{graphics}
\usepackage{amsmath}
\usepackage{siunitx}
\DeclareSIUnit\year{yr}
\DeclareSIUnit\evts{evts}
\DeclareSIUnit\NIU{NIU}
\usepackage[version=4]{mhchem}
\usepackage{epstopdf}
\usepackage{multirow}
\usepackage{url}
\usepackage{graphicx} 		
\usepackage{color}
\usepackage{enumitem}
\definecolor{ballblue}{rgb}{0.13, 0.67, 0.8}
\usepackage[
  pdftitle={AntineutrinosTheia},
  colorlinks=true,
  linkcolor=red,
  citecolor=red,
  menucolor=red,
  urlcolor=red
]{hyperref}
\usepackage[
    maxbibnames=99,
    maxnames=2,
    firstinits=true,
    uniquename=init,    
    uniquelist=false,
    backend=biber,
    doi=false,isbn=false,url=false,eprint=false,
    bibstyle=numeric,
    sorting=none
]{biblatex}
\renewbibmacro{in:}{}

\AtEveryBibitem{\clearfield{title}}

\newcommand{\Ep}{\text{E}_\text{prompt}}
\newcommand{\Ed}{\text{E}_\text{delay}}
\newcommand{\hl}{T_{1/2}}
\newcommand{\dt}{\Delta\text{t}}
\newcommand{\dr}{\Delta\text{R}}
\newcommand{\Theia}{\textsc{Theia}}
\newcommand{\ratpac}{RAT-PAC}

\usepackage[switch]{lineno}

\newcommand\blfootnote[1]{%
  \begingroup
  \renewcommand\thefootnote{}\footnote{#1}%
  \addtocounter{footnote}{-1}%
  \endgroup
}

\usepackage[compact]{titlesec}         
\titlespacing{\section}{0pt}{1em}{1em} 
\titlespacing{\subsection}{0pt}{1em}{1em} 
\titlespacing{\paragraph}{0pt}{1em}{1em} 
\titlespacing{\subparagraph}{0pt}{1em}{1em} 
\counterwithout{subparagraph}{paragraph}
\counterwithout{subparagraph}{subsubsection}
\counterwithout{subparagraph}{subsection}
\counterwithout{subparagraph}{section}
\counterwithout{paragraph}{subsubsection}
\counterwithout{paragraph}{subsection}
\counterwithout{paragraph}{section}
\begin{document}
\title{Geo- and reactor antineutrino sensitivity at THEIA}
\author{Stephane Zsoldos$^\dagger$\inst{1}\inst{2} \and Zara Bagdasarian\inst{1}\inst{2} \and Gabriel D. Orebi Gann\inst{1}\inst{2} \and Andrew Barna\inst{3} \and Stephen Dye\inst{3} 
}                     

%
\institute{
University of California, Berkeley, Department of Physics, CA 94720, Berkeley, USA \and 
Lawrence Berkeley National Laboratory, 1 Cyclotron Road, Berkeley, CA 94720-8153, USA \and
University of Hawai‘i at Manoa, Honolulu, Hawai‘i 96822, USA
}
\date{Received: date / Revised version: date}
%
\abstract{
We present the sensitivity of the \Theia{} experiment to low-energy geo- and reactor antineutrinos. 
For this study, we consider one of the possible proposed designs, a 17.8-ktonne fiducial volume \Theia{}-25 detector filled with water-based liquid scintillator placed at Sanford Underground Research Facility (SURF). 
We demonstrate \Theia{}’s sensitivity to measure the geo- and reactor antineutrinos via Inverse-Beta Decay interactions after one year of data taking with $11.9 \times 10^{32}$ free target protons. \newline
The expected number of detected geo- and reactor antineutrinos is $218\,^{+28}_{-20}$ and $170\,^{+24}_{-20}$, respectively.
The precision of the fitting procedure has been evaluated to be 6.72\% and 8.55\% for geo- and reactor antineutrinos, respectively. 
We also demonstrate the sensitivity towards fitting individual Th and U contributions, with best fit values of $N_\text{Th}=39^{+18}_{-15}$ and $N_\text{U}=180^{+26}_{-22}$. We obtain $(\ce{Th}/\ce{U})=4.3\pm2.6$ after one year of data taking, and within ten years, the relative precision of the $(\ce{Th}/\ce{U})$ mass ratio will be reduced to 15\%. \newline
Finally, from the fit results of individual \ce{Th} and \ce{U} contributions, we evaluate the mantle signal to be $S_\text{mantle} = 9.0\,\pm [4.2,4.5]$\,NIU.  
This was obtained assuming a full-range positive correlation ($\rho_c\in[0, 1]$) between \ce{Th} and \ce{U}, and the projected uncertainties on the crust contributions of 8.3\% (\ce{Th}) and 7.0\% (\ce{U}). 
When considering systematic uncertainties on the signal and background shape and fluxes, the mantle signal becomes $S_\text{mantle} = 9.3\,\pm [5.2,5.4]$\,NIU.
}
\maketitle
\blfootnote{$^\dagger$ Presently at King's College of London and the University of Tokyo Kavli Institute of Physics and Mathematics of the Universe.}
\section{Introduction}
    Antineutrinos were detected for the first time in 1956 by Clyde Cowan and Fred Reines by recording the transmutation of a free proton by particles born in nuclear reactors\,\cite{reines53}. This detection confirmed the existence of the neutrino and marked the advent of experimental neutrino physics. For decades, neutrino research has been an active and fruitful pursuit in the fields of particle physics, astrophysics, and cosmology. Neutrino experimental physics provided a glimpse into some of the most obscured astrophysical phenomena in the universe. The confirmed neutrino and/or antineutrino sources include the Sun\,\cite{Cleveland:1998nv}, nuclear reactors, particle accelerators\,\cite{K2K:2002icj}, the Earth\,\cite{Araki:2005qa,Borexino:2013tnm}, atmosphere\,\cite{Reines:1965qk}, and core-collapse supernovae\,\cite{Kamiokande-II:1987idp, Bionta:1987qt}. Moreover, the quantum mechanical phenomenon, known as neutrino oscillation, is our first observation beyond the Standard Model\,\cite{nobel_2015}, and it possibly holds the key to the explanation of the matter-dominated universe\,\cite{Bilenky:1978nj}. Since the neutrino discovery, reactor antineutrinos have continued to make huge contributions to studies of neutrino properties, including the measurement of neutrino oscillation parameters, neutrino mass ordering, and even the possibility of “sterile” neutrino flavors\,\cite{STEREO:2019ztb, PROSPECT:2020sxr}. Antineutrinos have also enabled a new interdisciplinary field of neutrino geoscience. Geoneutrinos are the only direct probe of radiogenic heat in the depths of the Earth, particularly the mantle, and can help discriminate between different geological models of Earth's formation and evolution. The number of electron flavour antineutrinos emitted in the radioactive decays of heat-producing elements (HPEs) with lifetimes compatible with the age of the Earth, such as \ce{^232Th}, \ce{^235U}, \ce{^238U}, and \ce{^40K}, is directly proportional to the HPE abundances and Earth’s radiogenic heat. 
    Borexino (Italy) and KamLAND (Japan) are the only two experiments to have observed geoneutrinos. The most recent measurements of geoneutrinos are $52.6^{+9.4}_{-8.6}$ (stat)$^{+2.7}_{-2.1}$ (sys) in Borexino\,\cite{Geo_Borexino} and $164^{+28}_{-25}$ in KamLAND\,\cite{Geo_KamLand,KamLAND_preliminary},
    corresponding to 18\% and 17\% precision. Expanding these measurements with higher statistics and to the other parts of the world will allow us to have a better understanding of radioactive elements abundances in the crust and mantle. SNO+\,\cite{snoplus} (Canada) and JUNO\,\cite{juno_2022} (China) will attempt geoneutrino measurements as part of their physics program in the near future. Of special interest is the concept of placing a neutrino detector on the seafloor as the oceanic crust is much thinner than the continental one, hence mantle contributions would dominate the measured geoneutrino flux.  With the original idea developed for Hanohano\,\cite{hanohano} to be placed at the oceanic floor near Hawaii, more recently, a collaboration in Japan reinvigorated this idea with the Ocean Bottom Detector (OBD)\,\cite{snowmass-OBD}. Other proposed experiments that hope to measure geoneutrinos are Jinping\,\cite{Jinping} in China, and the \Theia{} multipurpose detector\,\cite{theia_wp}. The latter is the focus of this paper. \Theia{} is a proposed large-scale scintillation-based neutrino detector that will deploy new target media, photon detectors, readout techniques and reconstruction algorithms to help discriminate between Cherenkov and scintillation signals. 
    The considered detector design consists of a cylindrical tank viewed by inward-looking photomultipliers (PMTs) and filled with water-based liquid scintillator (WbLS)\,\cite{wbls}, a mixture of water and an organic oil-based scintillator, combined using surfactants. 
    This novel target allows for the combination of directional sensitivity from the Cherenkov signal, with the low energy threshold and good resolution from the scintillation. Hence, \Theia{} has a broad physics program ranging from low energy solar to high energy accelerator neutrinos\,\cite{theia_wp}.

        \begin{figure}[!htb]
            \centering
            \resizebox{0.49\textwidth}{!}{
                \includegraphics{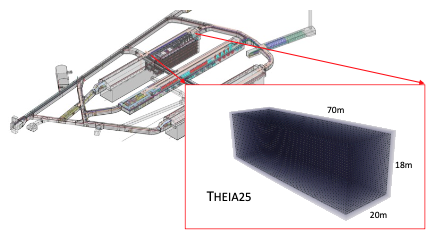}
            }
            \caption{
                A sketch of \Theia{}-25 sited in the planned
fourth DUNE cavern at SURF and a box-shape detector outline. 
                The maximum space available in terms of fiducial volume is shown.
            }
            \label{fig:Theia-25kT}
        \end{figure}

    In this paper, we consider a \Theia{} design that would fit in a cavern the size and shape of those intended for the Deep Underground Neutrino Experiment (DUNE) at Sanford Underground Research Facility (SURF), which we call \Theia{}-25.
    Figure~\ref{fig:Theia-25kT} shows a sketch of the available space on-site and a proposed design for \Theia{}-25, a letterbox detector of dimensions $\SI{70}{\meter}\times\SI{20}{\meter}\times\SI{18}{\meter}$. 
    
    We explore the sensitivity of \Theia{}-25 towards geoneutrinos based on simulations of a one-year data-taking equivalent. \Theia{}'s first high statistics geoneutrino measurement in North America will be complementary to measurements in Asia and in Europe. A combined analysis, with contributions from experiments across the globe, is critical for understanding the contributions of the crust and mantle. \Theia{}’s good energy resolution also offers the potential to extract the  \ce{Th}/\ce{U} mass ratio from a spectral fit. We also estimate the sensitivity of \Theia{} at SURF towards the  antineutrinos originating at nuclear reactors at baseline of more than 700 km.
    In Section~\ref{sec:sig_bkg}, we give a short overview of the antineutrino sources, the advantages of their detection, as well as the assumptions on the main background sources. We proceed by describing the full Monte Carlo simulations we have performed in Sec.~\ref{sec:mc}, including all the inputs. Section~\ref{sec:analysis} presents the analysis methods, while Sec.~\ref{sec:results} focuses on the results. 
        
\section{Signals and Backgrounds}
\label{sec:sig_bkg}
    \subsection{Antineutrino signals}

    The main natural antineutrino sources expected at \Theia{}-25 at SURF are radioactive elements in the crust and mantle of the Earth. Geoneutrinos are electron flavour antineutrinos emitted inside the Earth, in the radioactive decays of HPEs with lifetimes comparable with the age of the Earth (\SI[exponent-product=\ensuremath{\cdot}]{4.54e9}{\year}): 
    \ce{^232Th} ($\hl = \SI[exponent-product=\ensuremath{\cdot}]{1.40e10}{\year}$), 
    \ce{^238U} ($\hl = \SI[exponent-product=\ensuremath{\cdot}]{4.47e9}{\year}$), 
    \ce{^235U} ($\hl = \SI[exponent-product=\ensuremath{\cdot}]{7.04e8}{\year}$), 
    and \ce{^40K} ($\hl = \SI[exponent-product=\ensuremath{\cdot}]{1.25e9}{\year}$)\,\cite{IAEA}:
    \begin{align}
        \ce{^238U}        \rightarrow& \, \ce{^206Pb} + 8\alpha  + 6 e^{-} + 6 \bar{\nu}_e + \SI{51.7}{\mega\electronvolt} \nonumber\\
        \ce{^235U}        \rightarrow& \, \ce{^207Pb} + 7\alpha  + 4 e^{-} + 4 \bar{\nu}_e + \SI{46.4}{\mega\electronvolt} \nonumber\\
        \ce{^232Th}       \rightarrow& \, \ce{^208Pb} + 6\alpha  + 4 e^{-} + 4 \bar{\nu}_e + \SI{42.7}{\mega\electronvolt} \nonumber \\
        \ce{^40K}         \rightarrow& \, \ce{^40Ca}  + e^{-}    + \bar{\nu}_e + \SI{1.31}{\mega\electronvolt}~\mathrm{(89.3\%)} \nonumber\\
        \ce{^40K} + e^{-} \rightarrow& \, \ce{^40Ar}  + \nu _{e} + \SI{1.505}{\mega\electronvolt}~\mathrm{(10.7\%)}. \nonumber
    \end{align}
    As \ce{^40K} electronic capture produces neutrinos their detection is not discussed in this analysis.
    Geoneutrino measurements can shed light on abundances and distributions of radioactive elements inside the Earth beyond the reach of direct measurements by sampling. In each decay, the emitted radiogenic heat is in a well-known ratio to the number of emitted geoneutrinos, providing a way to directly assess the Earth’s heat budget\,\cite{Fiorentini:2007te}. 

In addition to antineutrinos produced from within the Earth, nuclear power plants produce abundant antineutrinos and are the strongest man-made source. Many nuclei, produced in the fission process of nuclear fuel, decay through $\beta$-processes with the consequent emission of electron antineutrinos with the energy up to \SI{10}{\mega\electronvolt}. 

        Antineutrinos can be detected via the inverse beta decay (IBD) reaction:
         \begin{equation}
            \bar{\nu}_{e} + p \rightarrow e^{+} + n,
            \label{feq:ibd_int}
        \end{equation}
        in which the free protons of hydrogen nuclei act as the target. 
        IBD is a charge-current interaction that proceeds only for electron flavoured antineutrinos. 
        Since the combined mass of the neutron and positron is greater than the mass of the proton, the IBD interaction has a kinematic threshold of \SI{1.806}{\mega\electronvolt}. 
             A positron and a neutron are emitted as reaction products in this process. 
        The positron promptly comes to rest and annihilates emitting two 511\,keV $\gamma$-rays from para-positronium and three $\gamma$-rays from ortho-positronium decays, yielding a ``prompt'' signal, with a visible energy $E_\text{vis}$, which is directly correlated with the incident antineutrino energy $E_{\bar{\nu}_e}$: 
        \begin{equation}
        E_\text{vis} \sim E_{\bar{\nu}_e} - \SI{0.784}{\mega\electronvolt}.
        \label{eq:Epro}
        \end{equation}
        The offset results mostly from the difference between the \SI{1.806}{\mega\electronvolt}, absorbed from $E_{\bar{\nu}_e}$ in order to make the IBD kinematically possible, and the \SI{1.022}{\mega\electronvolt} energy released during the positron annihilation.
        The emitted neutron initially retains information about the ${\bar{\nu}_e}$ direction. 
        However, the neutron is detected only indirectly, after it is thermalized and captured, mostly on a proton. 
        Such a capture leads to an emission of a \SI{2.22}{\mega\electronvolt} $\gamma$-ray, which interacts typically through several Compton scatterings and is detected in a delayed signal. 

    \subsection{Overview of background sources}
        The time and spatial coincidence between prompt and delayed signals offer a clean topology for $\bar{\nu}_e$ IBD interactions, which strongly suppresses backgrounds. 
        Nevertheless, there are some non-antineutrino backgrounds that can imitate the IBD signature. 
        The rates of these backgrounds depend on the selection cuts applied to the search of prompt and delayed event coincidences. Assuming the same neutron capture time in WbLS as in water of $\tau_\text{n}=\SI{202}{\micro\second}$\,\cite{SNO:2020bdq}, an upper-bound for the time search of correlated events for IBD pairs candidates of $\dt_\text{cut}\equiv\SI{1}{\milli\second}$ is defined, corresponding to $5\tau_\text{n}$. However, the final selection cuts on the space and time correlation have been optimized and will be discussed in Sec.~\ref{subs:box}. Backgrounds mimicking the IBD interaction can be divided into two categories: 1) two independent sources produce prompt- and delayed-like events, 2) a sole physical process produces both the prompt and delayed events correlated in space and time.
We will refer to these categories as accidental and correlated backgrounds, respectively. 
            
        In the following, we give a brief overview of the backgrounds that can potentially contribute to the antineutrino search and classify them into those that require Monte Carlo simulations and those that can be safely neglected for this study.
        
            \paragraph{Accidentals}
            Radioactive impurities inside the PMTs glass bulb or dissolved in the WbLS are the primary sources of accidental background. 
            The cleaner the PMT glass and target material, the less is the probability of accidental coincidence to satisfy the search for prompt and delayed event pairs. 
        All these contributions were simulated and are discussed in Sec.~\ref{subsec:acc}. 
            
            \paragraph{Correlated from cosmogenic backgrounds}
                A high-energy cosmic muon can produce a copious amount of activated radioisotopes while knocking off nuclei along its path. 
                Whilst in most cases these muons are easily recognizable as they leave a very clear signature inside the detector losing about \SI{2}{\mega\electronvolt\per\centi\meter}, the spallation products can be missed, creating an IBD-like signal from their subsequent decay. 
                Short-lived radioisotopes can be significantly suppressed by using a time veto cut. 
                However, long-lived radioisotopes can create a coincidental background long after the muon has been detected.
    
                A previous study using data from Super-Kamiokande and FLUKA simulations evaluated the amount of radioisotopes spallation products from the muon track in water \cite{Li:2014sea}.
                Assuming the same yield in WbLS, the expected spallation production rates inside \Theia{}-25 are evaluated using Equation~\ref{eq:spallation-rates}:
                \begin{eqnarray}
                    \phi_i &=& Y_i B_i P_i(\mu_\text{cut}) \cdot \Phi_\mu \label{eq:spallation-rates} \\
                           &=& Y_i B_i \exp{\left(-\ln{2}\,\frac{\mu_\text{cut}}{T_{{1/2}_i}}\right)} \cdot \int\limits_\theta \phi_\mu \mathcal{A} l_\mu \mathrm{d}\theta, \nonumber
                \end{eqnarray}
                where $Y_i$ is the yield of radioisotope $i$ taken from Ref.~\cite{Li:2014sea}, $B_i$ is the branching ratio for the considered decay, $P_i$ is the probability of the isotope with half-life $T_{{1/2}_i}$ to survive the muon veto cut $\mu_\text{cut}$, and $\Phi_\mu$ is the product of the muon flux and its path length inside the detector, integrated over all directions. The term $\Phi_\mu$ does not depend on the radioisotope $i$ and can be calculated with the average path length of all muons inside \Theia{}-25 $\bar{l}_\mu$, assuming an overall angular distribution of cosmic muons to be $\propto\cos^n\theta_\text{Z}$\,\cite{Shukla:2016nio}, the coverage area above \Theia{}-25 $\mathcal{A}$, and the integrated muon flux, measured previously at SURF to be $\phi_{\mu} = \SI{5.31e-9}{\mu\per\second\per\centi\square\meter}$\,\cite{Muon_flux_Majorana}.
          
                Spallation products of interest can be divided into two categories: ($\beta^-$, n) emitters creating a correlated background, and $\beta^\pm$ emitters which can decay in coincidence with another spallation product from the same muon, creating a background that is fundamentally accidental but somewhat correlated in time and space. Tables~\ref{tab:spallation_rates} and~\ref{tab:spallation_single_rates} show the summary of the ($\beta^-$, n) and $\beta^\pm$ longest-lived radioisotopes of interest. 
                \begin{table}[!htb]
                    \caption{Summary of ($\beta^-$, n) emitters spallation products of interest inside \Theia-25 and expected rates inside the whole volume before and after a \SI{1}{\second} muon veto cut.}
                    \centering
                    \begin{tabular}{p{0.8cm}p{0.8cm}p{0.8cm}p{1.2cm}p{1.2cm}p{1.4cm}}
                    \hline\noalign{\smallskip}
                       Isotope& $\hl$  & $B_i$ & $Y_i$ [$10^{-7}$  & $\phi_i$  & $\phi_\text{veto}$\\
                        & \SI{}{[\second]} & & \,\SI{}{\per\mu\per\gram\cdot\centi\square\meter}] & [\SI{}{\evts\per\year}] & [\SI{}{\evts\per\year}] \\
                        \noalign{\smallskip}\hline\noalign{\smallskip}
                         \ce{^{17}N}  & 4.173 & 1    & 0.59 & 317  & 269\\
                         \ce{^{16}C}  & 0.747 & 1    & 0.02 & 10.8 & 4.25\\
                         \ce{^{9}Li}  & 0.178 & 0.51 & 1.90 & 520  & 9.6\\
                         \ce{^{8}He}  & 0.119 & 0.16 & 0.23 & 19.78 & 0.06\\
                         \noalign{\smallskip}\hline
                        \end{tabular}
                    \label{tab:spallation_rates}
                \end{table}
                A dedicated study to evaluate the impact of a muon veto cut on the spallation rates has been performed.
                For correlated backgrounds, the contribution of \ce{^{9}Li} and \ce{^{8}He} can be strongly suppressed by setting a \SI{1}{\second} muon veto cut.
                Its impact on the rates is shown by comparing the values in the last two columns of Table~\ref{tab:spallation_rates}. 
                This time cut will result in a dead-time of the experiment of 6.7\%, which is applied to the signal and other background rates.
                \begin{table}[!htb]
                    \caption{Most abundant single emitting $\beta$ products of spallation after a \SI{1}{\second} muon veto cut.}
                    \centering
                    \begin{tabular}{llllll}
                        \hline\noalign{\smallskip}
                         Isotope & Decay & $\hl$ \SI{}{[\second]} & $B_i$ & {$Y_i$} [$10^{-7}$ & $\phi_\text{veto}$ \SI{}{[\hertz]}\\
                          & & & & \,\SI{}{\per\mu\per\gram\cdot\centi\square\meter}] & \\
                        \noalign{\smallskip}\hline\noalign{\smallskip}
                         \ce{^{16}N} & $\beta^-$ & 7.132   & 0.94 & 18  & \SI{1.85e-4}{}  \\
                         \ce{^{8}Li} & $\beta^-$ & 0.838  & 1    & 13  & \SI{6.85e-5}{}  \\
                         \ce{^{8}B}  & $\beta^+$ & 0.770   & 1    & 5.8 & \SI{2.84e-5}{}  \\
                         \ce{^{9}Li} & $\beta^-$ & 0.178  & 0.49 & 1.9 & \SI{2.28e-7}{}  \\
                          \noalign{\smallskip}\hline
                    \end{tabular}
                    \label{tab:spallation_single_rates}
                \end{table}
                The probability that a radioisotope decays within a specific time interval $\dt$ is $P = \exp{\left(-\dt\ln{2}/\hl\right)}$.
                Taking into account all combinations of decays between isotopes from Table~\ref{tab:spallation_single_rates}, the resulting expected number of coincidence is estimated to be less than \SI{9e-4}{} events per year for $\dt=\dt_\text{cut}$.
                Therefore, from all the spallation products of cosmogenic muons only the leading contribution of \ce{^{17}N} is considered.
                The details of the simulation are discussed in Sec.~\ref{subsec:cor}.
    
            \paragraph{Correlated from fast neutrons}
                A fast neutron (with an energy $\geq\SI{1}{\mega\electronvolt}$) interaction on a WbLS nucleus can imitate the IBD signature by producing a proton recoil, falsely identified as a prompt signal, and subsequently being thermalized and captured on a hydrogen hence producing a correlated delayed signal.
                This recoil proton from neutron elastic scattering can produce measurable ionization, described by the empirical Birk's law\,\cite{Birks:1951boa}, which is an empirical formula for the light yield per path length as a function of energy loss per path length for an ionizing particle. 
                Birk's constant, $k_B$, which characterizes this process, has not yet been measured in WbLS at the \SI{}{\mega\electronvolt} scale, but can be interpolated from measurement at BNL with a high-energy proton beam\,\cite{Bignell:2015oqa}.
                
                These fast neutrons are also a spallation product of high-energy cosmogenic muons passing through matter. 
                Therefore their rates are estimated using the same parent cosmogenic muon flux as in the previous section.
                The characteristic fast neutron thermalization time constant during which a proton recoil may create a prompt signal is \SI{5.3}{\micro\second} \cite{fn_thermalization}, and afterwards the neutron capture time is taken as $\tau_\text{n}=\SI{202}{\micro\second}$. 
                Using the previously set muon veto cut of \SI{1}{\second}, the identification of the parent muon will strongly suppress all fast neutrons produced inside \Theia{}-25. 
                Therefore all fast neutrons produced by visible muons are not considered throughout this analysis.
            
                However, muons missing the detector may produce fast neutrons in the surrounding rock, and these can reach \Theia{}-25 inner volume. 
                A previous study\,\cite{fast_neutrons_PhysRevD.73.053004} provides a parametrization of the expected fast neutron flux produced along the cosmic muons track at various underground laboratories ($\phi_n=\SI{5.39e-10}{\centi\meter^{-2}\second}^{-1}$ at SURF), the neutron multiplicity ($m_n=7.02$) and mean energy ($\langle E_n \rangle = \SI{98}{\mega\electronvolt}$), together with the fraction of neutrons detected with respect to the distance of the muon track.
                Typically, the neutron flux is attenuated by about two orders of magnitude at distances larger than \SI{3.5}{\meter} from the muon track; however, as much as 10\% remain at distances from 2 to \SI{2.5}{\meter}\,\cite{fast_neutrons_PhysRevD.73.053004}.
              
                Integrating from \Theia{}-25 edges, the muon-induced neutron flux in the rocks surrounding the detector yields a rate:
                \begin{equation}
                    m_n \phi_n \int\limits_0^\infty (2x(L+H)+4x^2)e^{-x}\mathrm{d}x = \SI[exponent-product=\ensuremath{\cdot}]{6.70e-3}{n\per\second},
                \end{equation}
                with $L=\SI{70}{\meter}$ \Theia{}-25 length, $H=\SI{18}{\meter}$ \Theia{}-25 width, and $x$ the distance from \Theia{}-25 edges.
                Therefore the fast neutrons contribution to the correlated background is considered, and its simulation is discussed in Sec.~\ref{subsec:fastn}. 
                
            \paragraph{Correlated from atmospheric neutrino neutral current interaction}
                Atmospheric neutrinos can interact by neutral current quasielastic nucleon knock-out (NCQE) process on \ce{^{16}O}:
                \begin{eqnarray}
                    \nu + \ce{^{16}O} &\rightarrow& \nu + n + \ce{^{15}O^*} \label{eq:NCQE} \\
                    \nu + \ce{^{16}O} &\rightarrow& \nu + p + \ce{^{15}N^*}.
                \end{eqnarray}
                This interaction becomes dominant for $E_\nu \geq \SI{200}{\mega\electronvolt}$ until \SI{1}{\giga\electronvolt} when neutral current inelastic process without nucleon knock-out, $\nu + \ce{^{16}O} \rightarrow \nu + \ce{^{16}O^*}$, overtake\,\cite{PhysRevLett.108.052505}.
                The process shown in Equation~\ref{eq:NCQE} is a background to a low energy antineutrino search through IBD, because the excited \ce{^{15}O^*} will immediately decay into \ce{^{15}O} emitting $\gamma$-rays from nuclear de-excitation. 
                Reference~\cite{PhysRevLett.108.052505} provides a theoretical treatment of the probabilities of occurrences of different excited states~\cite{PhysRevD.100.112009}, summarized in Table~\ref{tab:O15}. 
                The $(p_{1/2})^{-1}$ is the ground state of \ce{^{15}O} and therefore does not emit $\gamma$-ray. 
                The $(p_{3/2})^{-1}$ almost always emits one $\gamma$-ray with \SI{6.18}{\mega\electronvolt} energy. 
                The higher energy states $(s_{1/2})^{-1}$ and everything higher, referred to simply as ``others'', have a large branching ratio to nucleons or alpha particles, which can lead to secondary $\gamma$-ray emissions. 
              At the moment, there is neither data nor a theoretical prediction of $\gamma$-ray emission for the higher energy states covered by others.
                Further detailed descriptions on the treatment of these states are given in~\cite{PhysRevLett.108.052505,Folomeshkin:1976vj}.
                \begin{table}[!htb]
                    \caption{Probabilities of \ce{^{15}O^*} state occurrences\,\cite{PhysRevD.100.112009}. }
                    \centering
                    \begin{tabular}{lll}
                        \hline\noalign{\smallskip}
                        \ce{^{15}O^*} state & $P$ & $\gamma$ [\SI{}{\mega\electronvolt}] \\
                        \noalign{\smallskip}\hline\noalign{\smallskip}
                         $(p_{1/2})^{-1}$ & 0.158  & -- \\
                         $(p_{3/2})^{-1}$ & 0.3515 & \SI{6.18}{\mega\electronvolt}\\
                         $(s_{1/2})^{-1}$ & 0.1055 & $>$ \SI{8}{\mega\electronvolt} \\
                         others           & 0.385  & \\
                         \noalign{\smallskip}\hline
                    \end{tabular}
                    \label{tab:O15}
                \end{table}
        The number of atmospheric NCQE interactions can be estimated by taking the convolution of the atmospheric neutrino flux with the electron and muon neutrino neutral-current cross-section.
            The atmospheric neutrino flux at SURF is estimated from the modified ``DPMJET-III'' model\,\cite{Honda:2006qj}. 
            The oscillation probability is calculated with the Osc3++ framework\,\cite{Super-Kamiokande:2019gzr}. 
            Since atmospheric neutrinos generally experience a varying matter profile, and hence electron density changes as they travel through the Earth, they experience a variety of matter effects\,\cite{Wolfenstein:1977ue}. The calculation of oscillation probability in this analysis takes such variation on matter density into consideration, with a simplified version of the preliminary reference Earth model (PREM)\,\cite{Super-Kamiokande:2017yvm}.
            NCQE cross-sections tables are taken from the NEUT framework\,\cite{Hayato:2009zz}.
            For NCQE interactions the nominal nucleon momentum distribution is based on the Benhar spectral function\,\cite{Ankowski:2011ei,Benhar:2005dj}. The expected atmospheric neutrino rate is given by:
            \begin{eqnarray}
                N = \mathcal{N}_{n\ce{^{16}O}} \times 
                \int\limits_{\SI{200}{\mega\electronvolt}}^{\SI{1}{\giga\electronvolt}}
                \int\limits_{0^\circ}^{90^\circ} 
                \phi(E_\nu, \theta) P(E_\nu, \theta) \times \nonumber\\ 
                \sigma_\text{NCQE}(E_\nu)
                \text{d}E_\nu \text{d}\theta\nonumber,
           \end{eqnarray}
            \noindent with $E_\nu$ the neutrino energy, $\theta$ the zenith angle, $\mathcal{N}_{n\ce{^{16}O}}$ the number of neutron target available, $\phi(E_\nu, \theta) P(E_\nu, \theta)$ the neutrino oscillated flux, and $\sigma_\text{NCQE}(E_\nu)$ the neutrino-oxygen neutral-current quasi-elastic (NCQE) cross-section. 

            After integration, the expected atmospheric neutrino NCQE rate is $\phi_{{\text{atm}}}=\SI{3.25e-06}{\hertz}$. 
            The expected rate of excited $(p_{3/2})^{-1}$ \ce{^{15}O^*} rate will be the product of $\phi_{{\text{atm}}}$ with the respective branching ratio, yielding \SI{1.14e-6}{\hertz} or 36.0 events per year.
            Only the dominant contribution from the $(p_{3/2})^{-1}$ \ce{^{15}O^*} excited state is simulated and discussed in Sec.~\ref{subsec:NCQE}.

        \paragraph{(${\alpha}$, n) background}      
            
            Energetic ${\alpha}$ particles, generated in $\ce{Po}$ decays along the $\ce{U}$ and $\ce{Th}$ chains, as well as an out-of-equilibrium \ce{^210Po}, can produce neutrons by capture on certain isotopes contained within the detector, specifically \ce{^10B}, \ce{^11B}, \ce{^13C}, \ce{^17O}, \ce{^18O}, \ce{^29Si}, and \ce{^30Si}. 
            Assuming natural abundances of these isotopes in 3\% WbLS and the PMT's borosilicate glass, the typical mean energy of the neutron spectrum is $\langle E_N \rangle=\SI{3}{\mega\electronvolt}$.
            Although the average energy of these neutrons is softer compared to their cosmogenic cousins, they can imitate the IBD signal by producing a fast proton recoil followed by a neutron capture on hydrogen.
            
            Additional contributions can also arise in the process of $\ce{^18O}(\alpha,n)\ce{^21Ne^*}$ and its equivalent with \ce{^17O}.
            As \ce{^21Ne^*} decays by neutron emission, each reaction of this kind produces two neutrons.
            Recorded events with a multiplicity larger than two can be efficiently removed, however the possibility remains that the quenched proton recoil goes undetected, and both neutron captures create an IBD-like topology, with one being mistaken for a positron prompt signal and second one for the delayed event.

            Moreover, neutron produced during ${\alpha}$ interaction on \ce{^13C}, 
            \begin{equation}
        	    \ce{^13C} +\alpha \longrightarrow \: \ce{^16O}^* + n,
        	\end{equation}
            can satisfy the delayed event search, and there are three possibilities for the generation of the event that can imitate an IBD prompt event\,\cite{Geo_Borexino}:
            \begin{itemize}
        	    \item recoil proton appearing after the scattering of the fast neutron on proton; 
                \item $\gamma$-emission with energy of \SI{6.13}{\mega\electronvolt} or \SI{6.05}{\mega\electronvolt}, as a result of \ce{^16O^*} de-excitation;
        	    \item \SI{4.4}{\mega\electronvolt} $\gamma$-ray that is a product of the two-stage process: First, \ce{^12C} is excited into \ce{^12C^*} in an inelastic scattering off a fast neutron. 
        	    Then, \ce{^12C^*} transits to the ground state, accompanied by the $\gamma$ emission:
        	    \begin{align}
        		    n + \: ^{12}\text{C} &\longrightarrow \: ^{12}\text{C}^* + n, \\
        		    ^{12}\text{C}^* &\longrightarrow \: ^{12}\text{C} + \gamma \: (4.4\,\text{MeV}).
        		    \label{eq:inelastic_scattering}
        	    \end{align}
            \end{itemize}
            Neutron per decay yield and energy spectrum of all above-mentioned (${\alpha}$, n) processes in 3\% WbLS target material (see Sec.~\ref{subs:det}) and PMT borosilicate glass have been calculated using the NeuCBOT\,\cite{NeuCBOT} software. 
            We assume SNO cleanliness level for water\,\cite{SNO_water} (see Sec.~\ref{subsec:acc}), Borexino Phase-I cleanliness level for LS\,\cite{Geo_Borexino}, and isotope concentrations in the PMTs glass from Table~\ref{tab:accidentals_rates}. 
            All materials are simulated with their natural isotopes abundances.
            
            The total contributions expected from \ce{U} and \ce{Th} chains in WbLS are 2.16 on \ce{^13C}, 1.80 on \ce{^17O}, and 15.2 on \ce{^18O} events per year. 
            Only the dominant contribution from $\ce{^18O}(\alpha,n)\ce{^21Ne^*}$ is simulated and discussed in Sec.~\ref{subsec:AlphaN}.
            
            The total neutron rate expected from \ce{U} and \ce{Th} chains contained in the PMT glass is \SI{1.05}{\hertz}.
            Almost all of these fast neutrons come from the $\ce{^11B}(\alpha, n)\ce{^14C}$ interaction, mostly produced during the \ce{^238U} lower decay chain.
            The expected neutron spectrum is simulated and discussed in Sec.~\ref{subsec:AlphaN}.

        \paragraph{(${\gamma}$, n) background}  
             
            The only (${\gamma}$, n) reaction that can be triggered by \ce{^208Tl} $\gamma$-rays is the photo-dissociation from deuterium: 
            \begin{equation}
                \ce{^2H}+\gamma \rightarrow \ce{^1H}+n.
            \end{equation}
            The above reaction has a threshold of \SI{2.22}{\mega\electronvolt}, while reactions on various isotopes of carbon and oxygen have thresholds that range from \SI{4.10}{\mega\electronvolt} to \SI{18.7}{\mega\electronvolt}, well beyond the energy of $\gamma$-rays occurring from natural radioactivity.
            From the \ce{^208Tl} gamma spectrum end point, the mean neutron energy is expected to be significantly below \SI{1}{\mega\electronvolt}, therefore producing a very soft proton recoil spectrum, followed by a capture on hydrogen. 
            Nevertheless, the coincidence with the deposited gamma energy above threshold and the neutron capture can imitate the IBD signal.

            Using the fraction above the (${\gamma}$, n)\ce{^2H} threshold of the \ce{^208Tl} spectrum for both water and PMTs, the \ce{^2H} photo-dissociation cross-section integrated above the threshold\,\cite{gamma_n_cross_section}, the number of \ce{^2H} per gram of WbLS, and the $\gamma$-ray attenuation length in water, we obtain \SI{1.78e-2}{\hertz}, essentially produced at the edge of the detector.
            This process lies two orders of magnitude below the ($\alpha$,n) production on the PMTs' glass and would be negligible. 
            Furthermore, only a fraction of this rate would produce a correlated IBD-like event.
            Treating the other fraction of this rate as a single event producing a \SI{2.22}{\mega\electronvolt} $\gamma$-rays from the sole neutron capture, it is negligible compared to the rates of $\gamma$-rays from PMTs.
            Therefore, the (${\gamma}$, n)\ce{^2H} photo-dissociation can be safely dismissed.
             
\section{Monte Carlo simulation}
\label{sec:mc}

    \subsection{Detector configuration}
    \label{subs:det}
    
        The detector configuration is modeled as a right cylinder with \SI{18}{\meter} diameter and \SI{70}{\meter} height. 
        Even though a letterbox \Theia{}-25 detector would be deployed at SURF as shown in Fig.~\ref{fig:Theia-25kT}, the right cylinder geometry was chosen as it was readily available within the simulation framework. 
        Furthermore, the reconstruction algorithm described in Sec.~\ref{subs:recon} had been previously set up to work with this right cylinder geometry.
        Since this analysis relies on a volume fiducialization to optimize the signal over background ratio, edge effects that could affect event reconstruction at the letterbox's corners can be neglected, the reconstruction behaving similarly between both geometries far from the edges. 
        Additional information about the considered detector designed can be found in Ref.\,\cite{theia_wp}.
        
        The expected target volume of this configuration would be \SI{17.8}{\kilo\tonne} (corresponding to \num{11.9e32} free protons).
        Facing inwards arranged against the inner wall of the cylinder are located 79432 10" Hamamatsu R7081-100 PMTs, with 34\% quantum efficiency (QE) and \SI{1.5}{\nano\second} transit time spread (TTS). 
        This number of PMTs corresponds to an effective coverage of the detector walls and caps at 90\%, which is the highest photo-coverage achievable (in terms of packing capacity). 
        Lower coverage can be studied by scaling the number of PMT hits prior to the event reconstruction.       
        A \SI{3}{\kilo\hertz} dark rate is assumed for this PMT model operating at a nominal gain around 10$^\circ$C (in equilibrium with the water temperature, considering SURF depth). The threshold for trigger was defined as 8 PMT hits within \SI{600}{\nano\second}. 
        Once this condition is satisfied, the trigger time is defined as the time of the first hit, contributing to this cluster. 
        Simulation of the neutrino interactions and radioactive decays is performed using the Geant4-based\,\cite{geant4} \ratpac{} framework\,\cite{ratpac}. 
        Cherenkov photon production is handled by the default Geant4 model, G4Cerenkov. 
        Rayleigh scattering is implemented by the module developed by the SNO+ collaboration\,\cite{snoplus_private}. 
        GLG4Scint model handles the generation of scintillation light, as well as photon absorption and reemission. The total number of scintillation photons is not expected to change linearly with energy due to quenching. This is taken into account in the simulation with Birk’s law\,\cite{Birks:1951boa}, the deposited energy after quenching $E_q$ is $\frac{E_q}{dx}=\frac{dE/dx}{1+k_b dE/dx}$, where $k_b$ is Birk’s constant. 
        
        We chose 3\% WbLS (3\% liquid scintillator, 97\% water) as the baseline target material to simulate. The inputs to the optical model used in the simulation are primarily based on data and bench-top measurements. The light yield (502 scintillation photons/MeV), the scintillation emission spectrum and time profile with the risetime of \SI{0.265}{\nano\second} were interpolated from the 1\%, 5\%, 10\% WbLS measurements from \cite{chess_wbls} and \cite{drew_wbls}. Preliminary results show that absorption lengths are long and scattering is the dominant mode of light loss.
        Due to no published measurements available for the absorption length in this material, the model as described in \cite{PhysRevD.103.052004} was used in the simulation. The implemented scattering length ($\simeq\SI{40}{\meter}$ at \SI{430}{\nano\meter}) and refractive index (1.35 above \SI{400}{\nano\meter}) were measured as the function of the wavelength at BNL\,\cite{bnl_private}.

    \subsection{Signal simulation}
\begin{figure}[!htb]
            \centering
            \resizebox{0.49\textwidth}{!}{
              \includegraphics{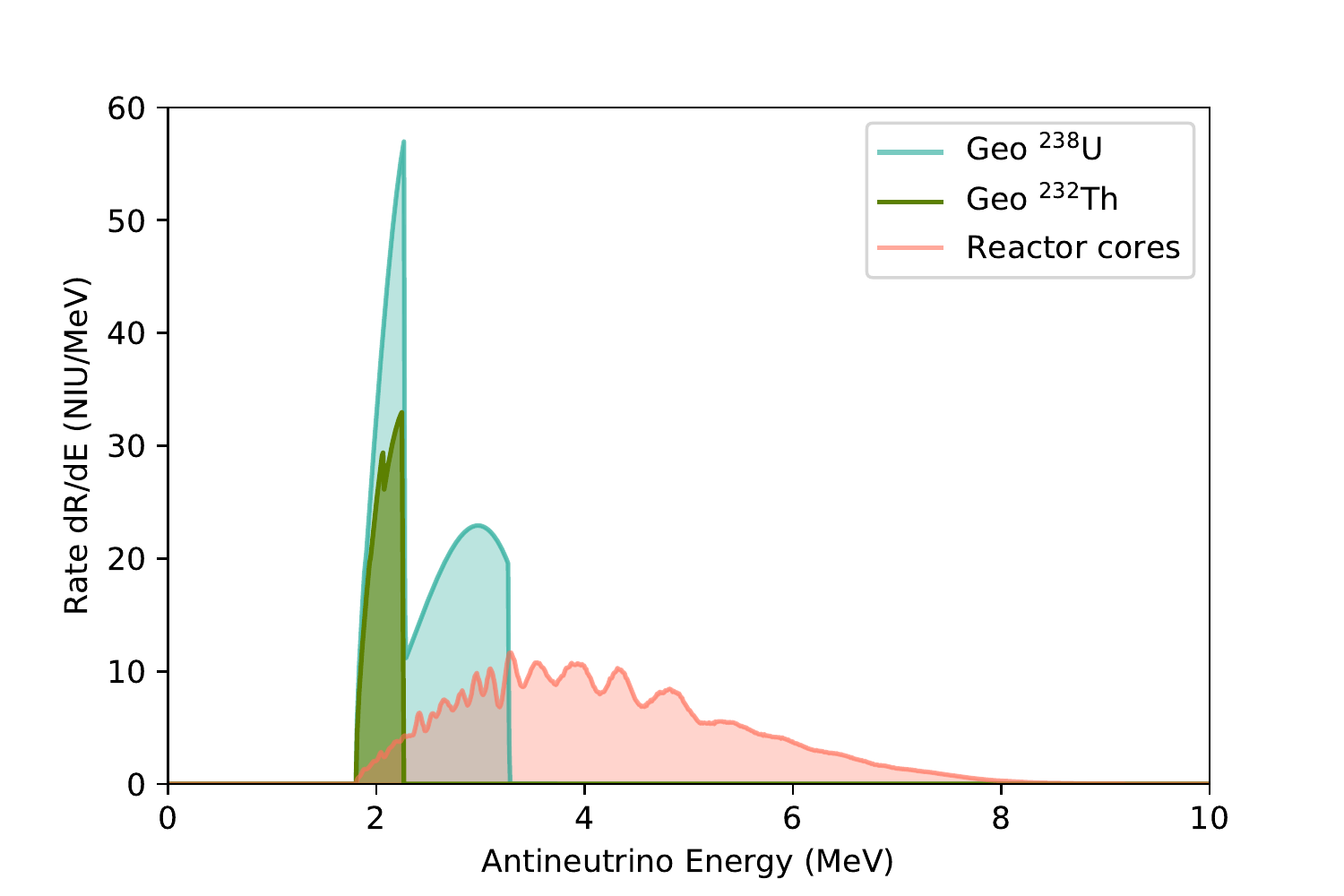}
            }
            \caption{The expected rate of antineutrino interactions at SURF, as the function of antineutrino energies. 1 NIU (Neutrino Interaction Unit) = 1 IBD interaction/$10^{32}$ targets/year. Source: geoneutrinos.org \cite{Dye:2015bsw}}
            \label{fig:input_spectra}       
        \end{figure}

        A dedicated generator inside the \ratpac{} software uses the positron energy spectrum as input to generate IBD pairs. 
        The positron spectrum is calculated from the expected antineutrino flux at SURF, available through the geoneutrino.org web app\,\cite{Dye:2015bsw}, uses decay spectra from \cite{Enomoto_Geo_Flux} and parameterized IBD cross section from \cite{Strumia:2003zx}. 
        Figure~\ref{fig:input_spectra} shows the antineutrino energy spectra used, with the (\ce{Th}/\ce{U}) ratio fixed to \num{4.33} for the crust and \num{3.9} for the mantle, while Table~\ref{tab:geo_inputs} summarizes the numerical values of the geoneutrino signal, with the breakdown between U and Th, as well as mantle and crust. 
             \begin{table}[!htb]
                \caption{The inputs for the geoneutrino signal simulation. 1 NIU (Neutrino Interaction Unit) = 1 IBD interaction/$10^{32}$ targets/year }
                \centering
                        \begin{tabular}{llll}
                        \hline\noalign{\smallskip}
                     &  S(U) [NIU] & S(Th) [NIU]  & S(Th)/S(U)\\ 
                    Crust    & 28.6 & 8.5 & 0.297\\
                    Mantle  & 7.3 & 2.0 & 0.274 \\
                    Total    & 35.8  & 10.5 & 0.293 \\                    
                    \noalign{\smallskip}\hline
                \end{tabular}
                \label{tab:geo_inputs}
            \end{table}
            
        For the reactor neutrinos spectrum, the nominal thermal power and the monthly load factors originate from the power reactor information system (PRIS), developed and maintained by the International Atomic Energy Agency (IAEA)\,\cite{PhysRevD.91.065002, ferrara_reactors}.
        The given spectra already consider the neutrino oscillation effects, with the values of oscillation parameters taken from NuFit v5.0, specifically the global analysis excluding the Super-Kamiokande atmospheric neutrino data\,\cite{Esteban:2020cvm}.

        IBD interactions are generated throughout the total WbLS volume available in \Theia{}-25.
        The number of expected IBD interactions from reactor and geoneutrinos can be found in Table~\ref{tab:SURF_rates}.
         We also show the geoneutrino signal breakdown in crust and mantle contributions, expected at SURF.

       \begin{table}[!htb]
            \caption{Antineutrino interaction rates in one year in the full \Theia{}-25\,kT volume.}
            \centering
            \begin{tabular}{ccccc}
                \hline\noalign{\smallskip}
                \multirow{4}{*}{Interactions [\SI{}{\evts\per\year}]}& \multicolumn{2}{c}{$R_\text{rea}$} & \multicolumn{2}{c}{$R_\text{geo}$} \\
                & \multicolumn{2}{c}{384.7} & \multicolumn{2}{c}{551.4}  \\ \cline{4-5}
                & \multicolumn{2}{c}{} & $R_\text{crust}$ & $R_\text{mantle}$ \\
                & \multicolumn{2}{c}{} & 441.8 & 109.6 \\
                \noalign{\smallskip}\hline
            \end{tabular}
            \label{tab:SURF_rates}
        \end{table}

    \subsection{Background simulation}
       \paragraph{Accidentals}
           \label{subsec:acc}
            The accidental coincidence rate is estimated from the R7081-100 PMT glass activity measurements by the WATCHMAN collaboration\,\cite{watchman_private} and the \ce{U}- and \ce{Th}-chain
backgrounds in the target material to be at the level of SNO water\,\cite{SNO_water}.
            Table~\ref{tab:accidentals_rates} presents the concentration and the activity of each isotope used in the simulations, along with the calculated decay rate.
            The PMT glass mass is taken as \SI{1400}{\gram}.
            \begin{table}[!htb]
                \caption{Assumed isotope concentrations, based on the R7081-100 PMT glass activity measurements by the WATCHMAN collaboration \cite{watchman_private} and the SNO cleanliness level water activity \cite{SNO_water}, followed by the corresponding expected rates inside the full \Theia{}-25 volume.}
                \centering
                        \begin{tabular}{llll}
                        \hline\noalign{\smallskip}
                     &  Concentration & Activity & Rates \\ 
                     &  &   [\SI{}{\becquerel\per\gram}] &  [\SI{}{\hertz}]\\
                     \noalign{\smallskip}\hline\noalign{\smallskip}

                    PMT \ce{^{238}U}    & 0.043 [ppm]                 & \SI{1.24e4}{} & \SI{60.8e3}{} \\
                    PMT \ce{^{232}Th}   & 0.134  [ppm]                & \SI{1.45e3}{} & \SI{22.1e3}{} \\
                    PMT \ce{^{40}K}     & 0.004212 [ppm]              & \SI{2.59e5}{} & \SI{124.3e3}{} \\                    
                    Water \ce{^{238}U}  & \SI{6.6e-15}{[\gram/\gram]} & \SI{1.24e4}{} & \SI{1.46}{} \\                    
                    Water \ce{^{232}Th} & \SI{8.8e-16}{[\gram/\gram]} & \SI{1.45e3}{} & \SI{2.27e-2}{} \\  
                    \noalign{\smallskip}\hline
                \end{tabular}
                \label{tab:accidentals_rates}
            \end{table}
             A custom decay chain generator within \ratpac{} is used in order to simulate the spectra for these radioisotopes, starting from \ce{^{214}Bi} and \ce{^{208}Tl} for \ce{^{238}U} and \ce{^{232}Th} respectively. In the following, we will refer to these events with \ce{^{214}Bi} and \ce{^{208}Tl} designations.
            The \ce{^234Pa} contribution for \ce{^{238}U} and \ce{^228Ac}, \ce{^212Bi} for \ce{^{232}Th} are ignored since their lower rates would have a negligible impact on the expected rates. 
            The $\beta$ spectrum of \ce{^{40}K} is used directly.
            About $\mathcal{O}(10^6)$ events are simulated for each contribution. Contributions to the accidentals rate from any other single backgrounds can be safely neglected.
\paragraph{Correlated from cosmogenic backgrounds}
        \label{subsec:cor}
            The \ce{^{17}N} $\beta$ spectrum is simulated using \ratpac{}. 
            The delayed neutron is simulated separately as a \SI{2.22}{\mega\electronvolt}  $\gamma$-ray, and both events are reassembled as a correlated background during the creation of the dataset described at Sec.~\ref{sec:datasets}.
            About $\mathcal{O}(10^5)$ \ce{^{17}N} $\beta$ and \SI{2.22}{\mega\electronvolt} $\gamma$-ray events are simulated.
    
\paragraph{Correlated from fast neutrons}
          \label{subsec:fastn}
            Birk's constant for WbLS 3\% has been extrapolated from Ref.~\cite{Bignell:2015oqa} to be \SI{0.43}{\milli\meter\per\mega\electronvolt}.
            The fast neutron energy spectrum is taken as an exponential law with a mean value of $\langle E_n \rangle = \SI{98}{\mega\electronvolt}$ \cite{fast_neutrons_PhysRevD.73.053004} and simulated inside \ratpac{} at the edges of \Theia{}-25.
            About $\mathcal{O}(10^5)$ fast neutrons are simulated.
   
        \paragraph{Correlated from atmospheric NCQE interaction}
       \label{subsec:NCQE}
       The $(p_{3/2})^{-1}$ deexcitation state of \ce{^15O^*} is simulated using \ratpac{} as a \SI{6.18}{\mega\electronvolt} $\gamma$-ray.
        The delayed neutron is simulated separately as a \SI{2.22}{\mega\electronvolt} $\gamma$-ray, and both events are reassembled as a correlated background during the creation of the dataset described in Sec.~\ref{sec:datasets}.
        About $\mathcal{O}(10^5)$ \SI{6.18}{\mega\electronvolt} $\gamma$-rays from \ce{^15O^*} and \SI{2.22}{\mega\electronvolt} $\gamma$-rays are simulated.

        \paragraph{(${\alpha}$, n) background} 
        \label{subsec:AlphaN}
        
        The total neutron spectrum expected from the PMTs contaminants evaluated using the NeuCBOT software is simulated using \ratpac{} directly from the PMTs' glass.
        About $\mathcal{O}(10^6)$ events are simulated.
        
        $(\alpha, n)$ interaction in \Theia{}-25 yields a neutron energy spectrum peaked at \SI{3}{\mega\electronvolt}, which mostly produces single triggers inside 3\% WbLS (assuming previous value of Birk's constant). 
        The neutron yield coming directly from the fiducial volume of \Theia{}-25 is essentially produced by $\ce{^18O}(\alpha,n)\ce{^21Ne^*}$ as described in Sec.~\ref{sec:sig_bkg}. 
        Being conservative we consider that both proton recoil signals are under the trigger threshold, but the two successive neutron capture on hydrogen are visible. 
        Hence, the interaction is simulated using \ratpac{} as two successive \SI{2.22}{\mega\electronvolt} $\gamma$-rays.
        Both events are reassembled as a correlated background during the creation of the dataset described in Sec.~\ref{sec:datasets}.
        About $\mathcal{O}(10^5)$ \SI{2.22}{\mega\electronvolt} $\gamma$-rays are simulated.
            
\section{Analysis}
\label{sec:analysis}

\subsection{Methodology for analysis}
    In order to obtain the sensitivity of the \Theia{}-25 detector to antineutrinos, the following analysis steps were implemented:
        \begin{enumerate}
            \item Reconstruction of the event vertex position based on the PMT hit times.
            \item Creation of merged dataset using an iterative procedure to prune background and select candidate IBD pairs from all signal and background events. 
            \item Optimization of Region-of-Interest (ROI) to increase signal-to-background ratio for a given antineutrino signal.
            \item Sensitivity analysis based on the creation and spectral fit of one-year data equivalent toy experiments.
        \end{enumerate}
        
    \subsection{Reconstruction}
    \label{subs:recon}
    
        A dedicated code was developed to reconstruct event vertex positions based on the time-of-flight. We build the binned distribution of the PMT hits residuals, i.e., the difference between the PMT hit times $T_\text{Hit}$ and the time traveled by photons from the vertex interaction to the PMT position $\vec x_\text{Hit}$. 
        In the case of bin $i$ and corresponding bin width $\mathrm{dt}$, the time residual associated to bin $i$ is:
        \begin{equation}
            T_\text{Res}^i = \int\limits_\text{i}^\text{i+1} 
            \sum_\text{Hit} T_\text{Hit}-T_\text{est} - \frac{ \vec x_\text{Hit} - \vec x_\text{est} }{ c_\text{water} } ~\mathrm{dt},
        \end{equation}
        where $c_\text{water}$ is the speed of light in water. The fit works by performing a maximum negative log-likelihood (NLL) search of the estimated vertex time $T_\text{est}$ and position $\vec x_\text{est}$ from a probability density function (PDF), created from a simulation of \SI{5}{\mega\electronvolt} electrons at the center of the detector. 
        A seeding algorithm is based on multilateration -- similar to the GPS location algorithm -- of all PMT hits. Solving the set of time-of-flight equations, we create a cloud of vertices eligible as hypotheses. This procedure yields $\mathcal{O}(10^3)$ seeds to reconstruct.

\begin{figure*}[!hbt]
        \centering
            \resizebox{0.99\textwidth}{!}{
                \includegraphics{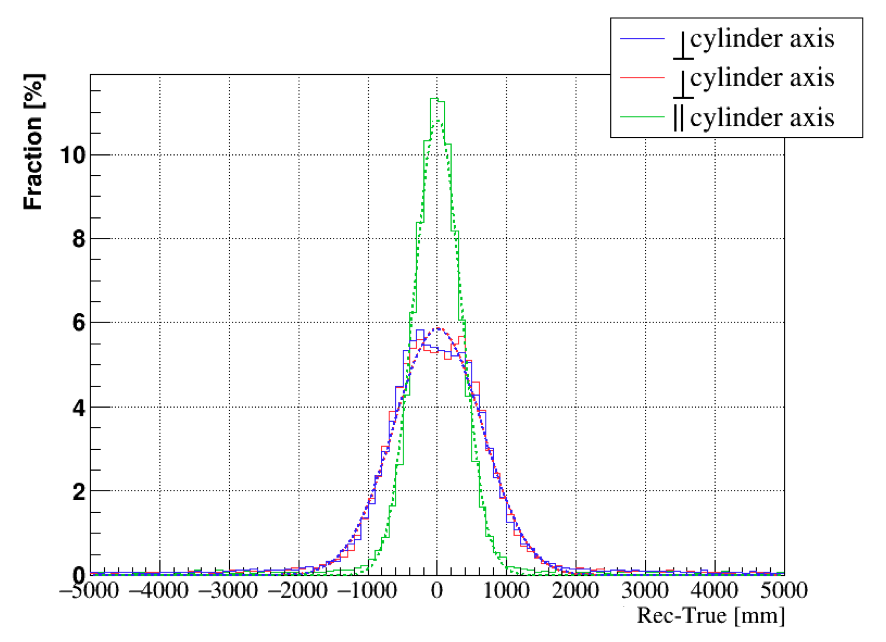}
                \includegraphics{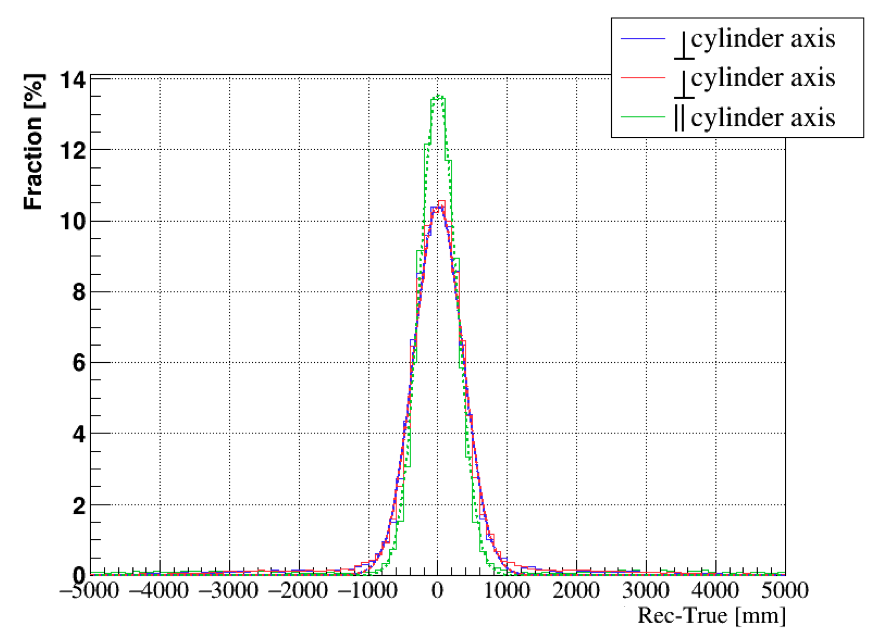}
          }
            \caption{Reconstruction performance with respect to positron true position, for geo- (left) and reactor $\overline{\nu}_e$ (right), integrated throughout the whole detector and energy spectrum.}
            \label{fig:reconstruction}       
        \end{figure*}

\begin{figure}[!hbt]
       \centering
            \resizebox{0.49\textwidth}{!}{
              \includegraphics{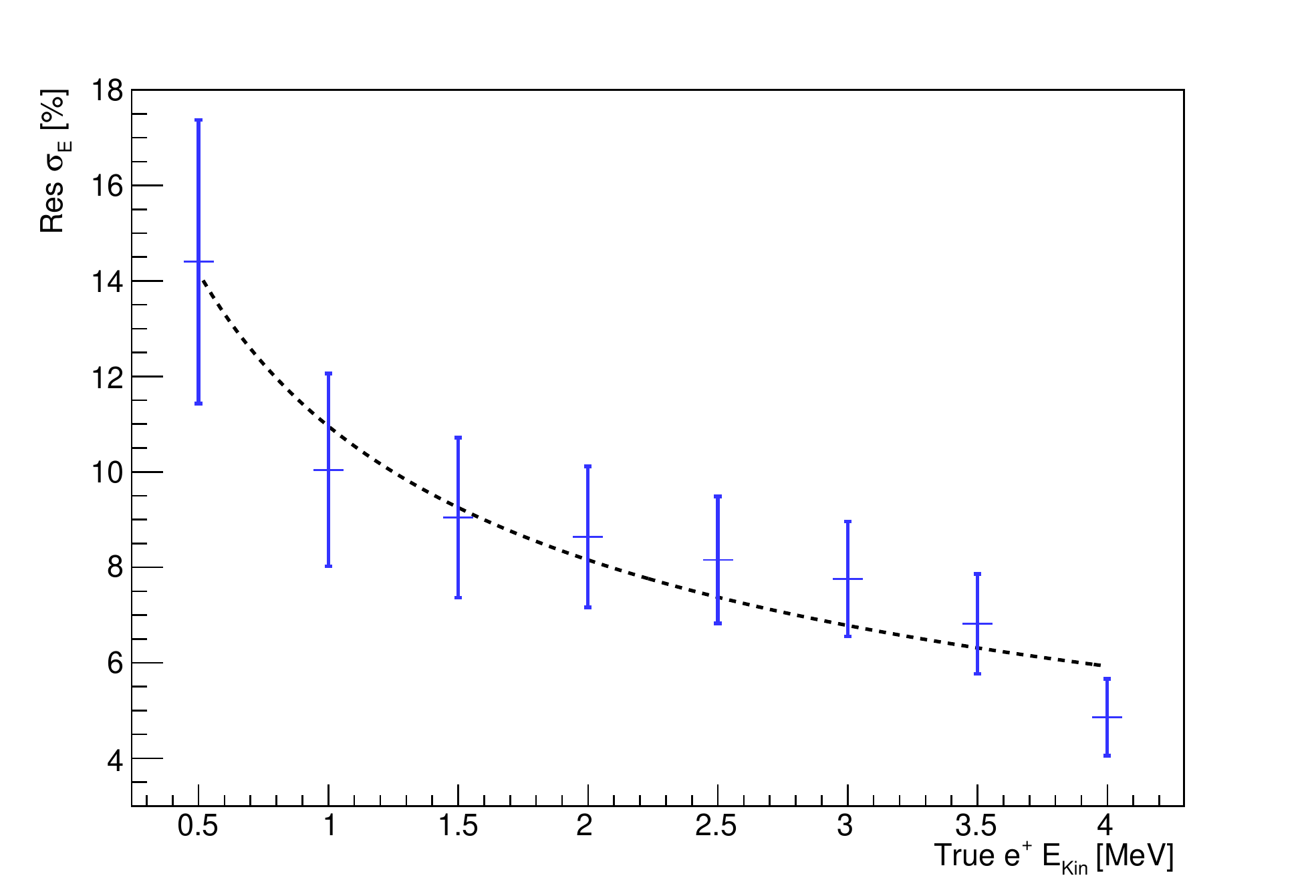}
            }
            \caption{Energy resolution for positrons generated at the center of the detector volume between \SI{0.5}{\mega\electronvolt} to \SI{4}{\mega\electronvolt}.}
            \label{fig:Energy_resolution}       
        \end{figure}

        Figure~\ref{fig:reconstruction} shows the performance of positrons reconstruction from IBD interactions simulated throughout the whole \Theia{}-25 volume. 
        For positrons from geoneutrino interactions vertex resolutions of ($58.0\pm0.3)\,\SI{}{\centi\meter}$ perpendicular to the cylinder axis and ($32.8\pm0.2)\,\SI{}{\centi\meter}$ parallel to the cylinder axis are achieved, while for reactor antineutrinos overall resolutions of ($34.3\pm0.2)\,\SI{}{\centi\meter}$ perpendicular to the cylinder axis and ($26.6\pm0.1)\,\SI{}{\centi\meter}$ parallel to the cylinder axis are obtained. 
        Moreover, we have mapped the energy resolution of the positrons between \SI{0.5}{\mega\electronvolt} and \SI{4}{\mega\electronvolt} generated at the center of the detector volume. 
        As shown in Fig.~\ref{fig:Energy_resolution}, the obtained energy resolution can be approximated as ${12\%}/{\sqrt{E}}$ for this particular choice of target material and detector configuration.
        
    \subsection{Event pruning and pairs dataset creation}
    \label{sec:datasets}

        The search for candidate pairs in the simulated data is dealt with by creating a single merged dataset of all signal and background contributions, which intersperses events in time in order to  mimic data from the detector.  
        This allows a full study of accidentals, including all correlations, as well as true coincidence backgrounds.
        
        The events are merged thanks to an iterative process.
        For each component $B_i$, an interval $\dt(B_i)$ is calculated:
        \begin{equation}
            \dt(B_i) = \frac{-\ln(1-u)}{\phi(B_i)},
        \end{equation}
        \noindent where $u$ is a random number generated from a uniform distribution between (0, 1) and $\phi(B_i)$ is the rate associated with component $B_i$.
        The shortest $\dt(B_i)$ is selected as the next event to be merged, and the time of the event is added to a previous point in time, starting at $t_0$:
        \begin{equation}
            t_{i+1} = t_i + \min_i{\left(\dt(B_i)\right)}.
        \end{equation}
        \noindent The procedure is iterated until $t_i =$ 1 year. 

        Special care is taken for some correlated backgrounds that do not have a dedicated Geant4 generator implemented in \ratpac{}, and were simulated in two stages: first, the equivalent prompt signal, and then the \SI{2.22}{\mega\electronvolt} $\gamma$-ray from the neutron capture on hydrogen. 
        To ensure the spatial correlation between the two events, both the equivalent prompt signal and \SI{2.22}{\mega\electronvolt} $\gamma$-ray were simulated at the center of the detector. 
        This spatial correlation is preserved during the merging procedure, when the prompt and delayed events pair is placed together at a random position within the detector volume.
        Lastly, each prompt event is assigned a corresponding delayed event with $\dt$ according to the neutron capture time, randomly sampling an exponential distribution $\exp{-\dt/\tau_{\text{n}}}$.
        \begin{figure*}[!htb]
            \centering
            \resizebox{0.99\textwidth}{!}{
                \includegraphics{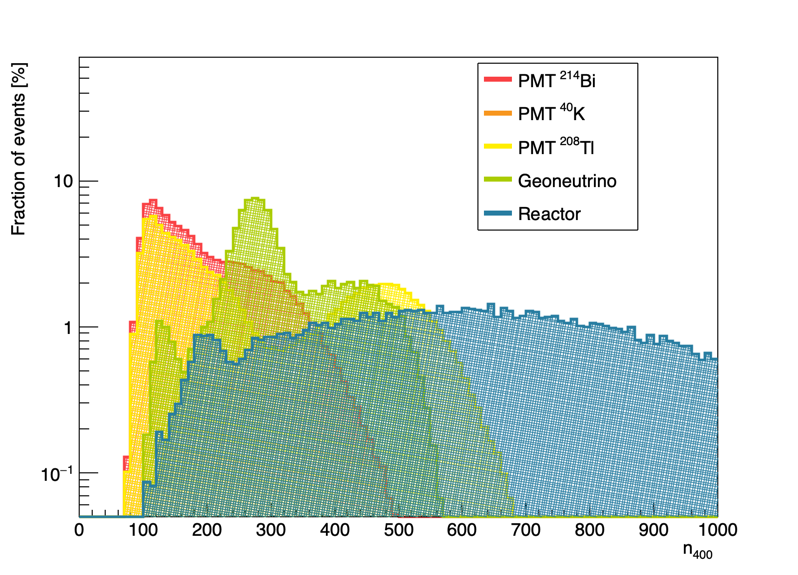}
                \includegraphics{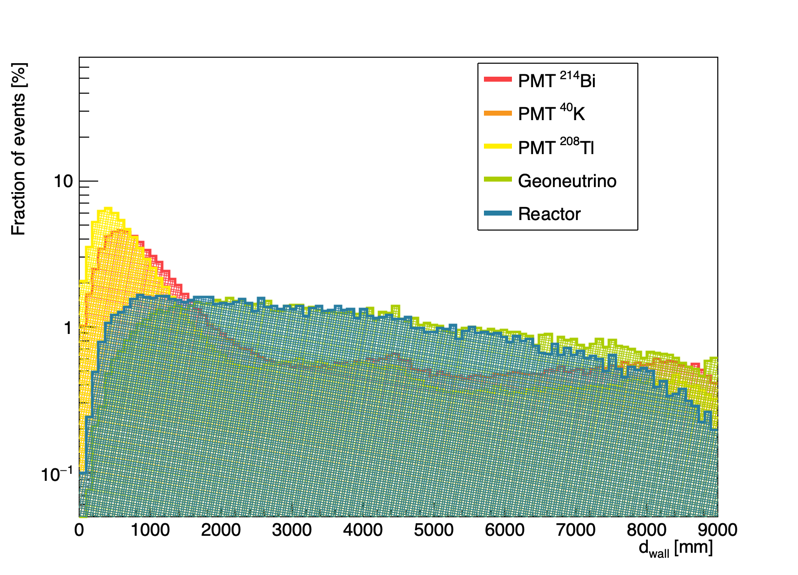}
                \includegraphics{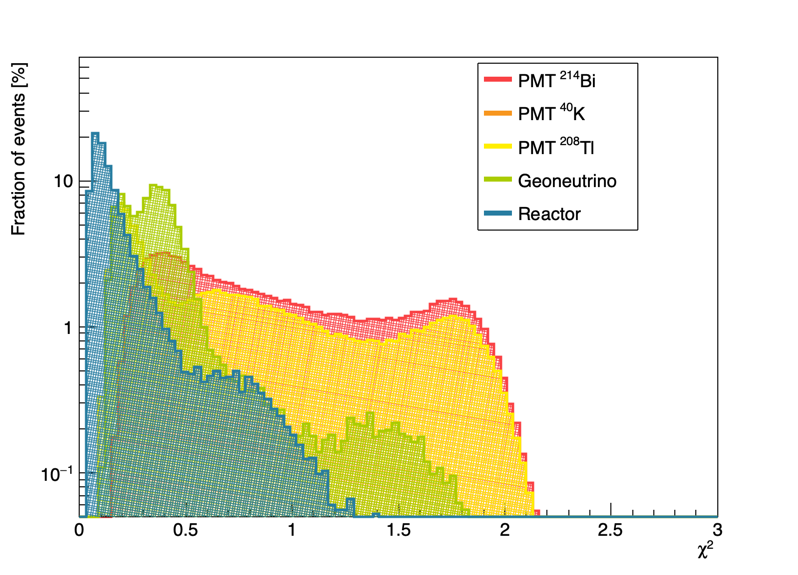}
            }
            \caption{PMT and signal event distributions of the characteristic parameters used in the BDT training: $n_{400}$ energy estimator (left), $d_\text{wall}$ distance (in \SI{}{\milli\meter}) of the reconstructed event from the closest wall (middle), and $\chi^2$ of the reconstruction fit (right). 
            }
            \label{fig:BDT_features} 
        \end{figure*}
        \paragraph*{}
        The high fission rates of radioisotopes contained in the PMT glass (see Table~\ref{tab:accidentals_rates}) will yield about $\mathcal{O}(10^5\SI{}{\hertz})$ background events during this iterative process. 
        Hence, extra steps are needed to prune the number of IBD candidates before saving the merged dataset to disk and performing the subsequent analysis steps.
        First, a cut on $\dt(B_i)\geq\SI{1}{\milli\second}$ discards events separated by more than $5\tau_n$. 
        Second, a cut $\dr\geq\SI{3}{\meter}$ between the distance of two consecutive events,
        \begin{equation}
            \dr = \vec{\mathbf{v}}_{B_{i+1}} - \vec{\mathbf{v}}_{B_{i}},
        \end{equation}
        discards events unlikely to originate from a single source. 
        Even considering the finite reconstruction resolution of the detector (see Fig.~\ref{fig:reconstruction}) and the neutron displacement during its thermalization, it is expected that the prompt and delayed events of a true IBD interaction are contained within $\dr$.
        
\begin{figure*}[!htb]
            \centering
            \resizebox{0.99\textwidth}{!}{
                \includegraphics{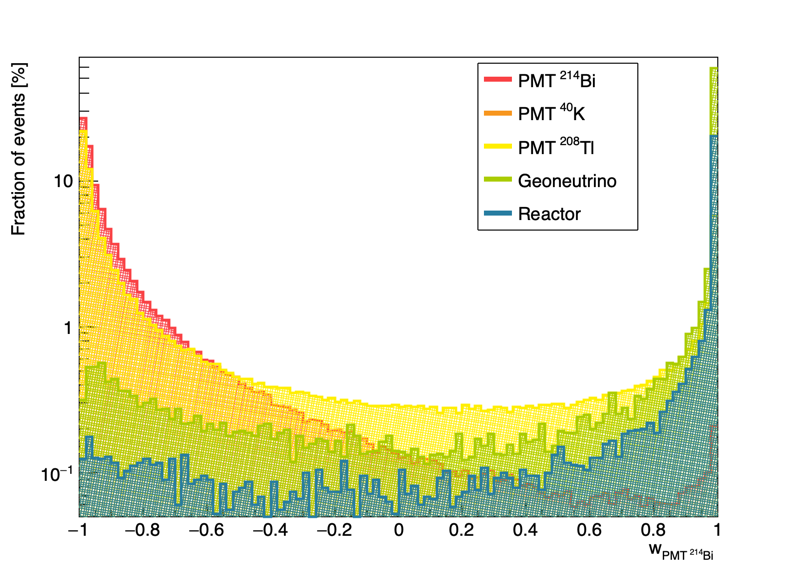}
                \includegraphics{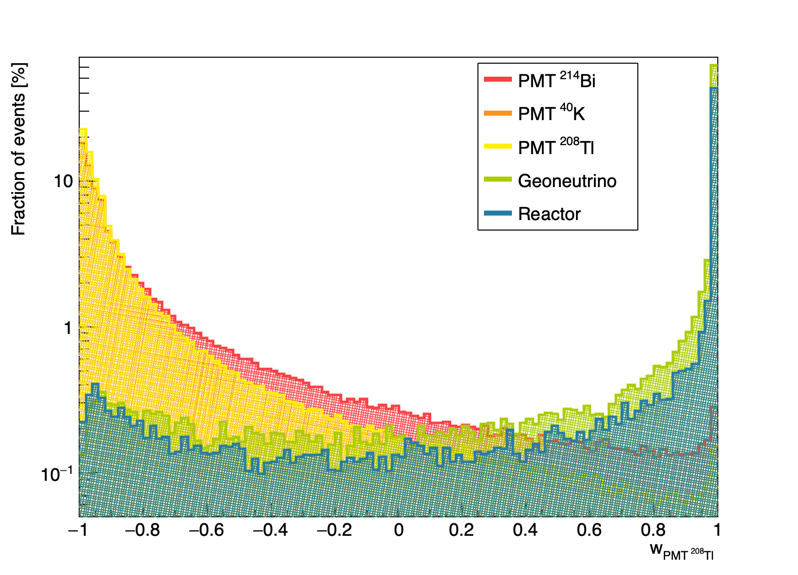}
                \includegraphics{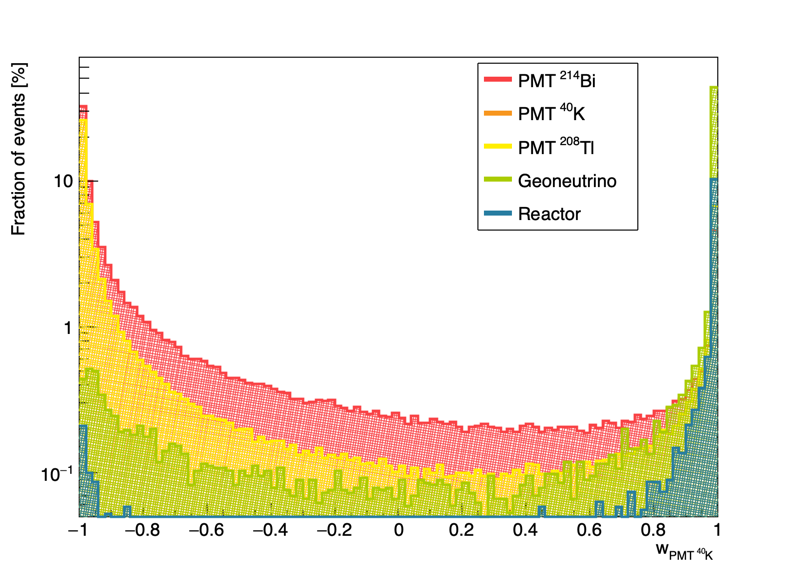}
            }
            \caption{BDT discriminant variable distributions for PMTs (in red) and signal (in blue) in case of \ce{^{214}Bi} $\gamma$-ray hypothesis (left), \ce{^{208}Tl} $\gamma$-ray hypothesis (middle), and \ce{^{40}K} $\gamma$-ray hypothesis (right).
            }
            \label{fig:BDT_results} 
        \end{figure*}

        Nevertheless, both $\dt$ and $\dr$ cuts do not yield enough suppression for the PMT accidental backgrounds. 
        Therefore, $\dt$ and $\dr$ cuts are complemented with an additional cut from a Boosted Decision Tree (BDT) in order to distinguish PMT radioisotope decays from the geo- and reactor antineutrinos prompt signal.
        A dedicated study has been performed to evaluate which features have the most powerful handle to discriminate this background from the signal.
        Figure~\ref{fig:BDT_features} shows three suitable parameters: an energy estimator of the event, $n_{400}$, which is the number of PMT hits in a \SI{400}{\nano\second} window after the trigger time; the shortest distance from a simulated detector wall to the reconstructed vertex, $d_\text{wall}$; and the Goodness-of-Fit $\chi^2$ of the reconstructed vertex, taken as the NLL of the hit time residuals distribution.
        A good fit from a radioisotope PMT $\gamma$-ray event would have a low NLL and be close to a detector wall, whereas a bad fit would have higher NLL and be further away from the wall.
        Signal events are distributed throughout the detector and a good reconstruction algorithm should yield no dependency between NLL and the true signal position at first order.
        Furthermore, it is expected that the fit quality must depend on the number of hits detected.
        As the number of hits $n_{400}$ can also be used to estimate the energy of the event, adding this feature to the BDT provides an additional handle from the $d_\text{wall}$ and $\chi^2$, which both depend on the reconstructed vertex position.
        The training of the BDT is performed, using CERN's ROOT library TMVA\,\cite{Hocker:2007ht}, against the hypothesis of being a signal, either a geo- or a reactor prompt event, or a $\gamma$-ray emitted by a \ce{^{214}Bi}, \ce{^{208}Tl} or \ce{^{40}K}. The following inputs are used:
        \begin{itemize}
            \item a total of \num{10000} PMT fission events for each decay chain, \ce{^{214}Bi}, \ce{^{208}Tl} and \ce{^{40}K};
            \item a total of \num{10000} IBD interaction events for geo- and reactor antineutrinos. Only the prompt signal is used in the training; 
            \item an additional \num{10000} \ce{^{214}Bi} and \ce{^{208}Tl} decay chains, corresponding to radioisotopes dissolved in water, are used as a control sample. 
            Since these \ce{^{214}Bi} and \ce{^{208}Tl} decay chains are similar to the PMTs events, but distributed uniformly throughout the detector, they can indicate if the BDT parameters are over-fitting the event classification.
        \end{itemize}
        The BDT yields a score $w_i$ representing a score under each background hypothesis to originate from radioisotope $i$: a signal event would tend to have a score of $1$, whereas a background-like event would tend towards $-1$.
        Figure~\ref{fig:BDT_results} shows the score for each background hypothesis, for all PMT $\gamma$-ray events and prompt signals from geo- and reactor antineutrinos.
        Using these scores, a BDT cut has been arbitrarily chosen to discriminate most background while preserving as much signal as possible:
        \begin{equation}
            \tau_{\text{BDT}} \equiv \min{\left( w_{\ce{^{214}Bi}}, w_{\ce{^{208}Tl}}, w_{\ce{^{40}K}} \right)} \geq 0.999.
        \end{equation}
        We apply BDT scoring to each individual background and signal event at the merging stage. The first column in Table~\ref{tab:eff_rates} provides the observed suppression factor by comparing the equivalent rate of each contribution before $\dt$ and $\dr$ cuts to the expected contribution calculated in Sec.\,\ref{sec:sig_bkg}.
        In addition, the contributions to the merged dataset pairs, which satisfy pre-optimized coincidence cuts ($\dt=\SI{1}{\milli\second}$ and $\dr=\SI{3}{\meter}$), and contributions of each component in the Region-of-Interest (ROI), optimized for geo- and reactor neutrinos signal, described in the following section can be also found in Table~\ref{tab:eff_rates}.
        \begin{table}[!htb]
        \centering
        \caption{
            The summary of all background and signal contributions at different analysis stages: the resulting post-BDT singles rates of each background inferred from the merged dataset, contributions to the merged dataset pairs, which satisfy pre-optimized selection cuts ($\dt=\SI{1}{\milli\second}$ and $\dr=\SI{3}{\meter}$), and contributions of each component in the Region-of-Interest (ROI), optimized for geo- and reactor neutrinos signal, described in the following section.
            }
                        \label{tab:eff_rates}
           \begin{tabular}{
          >{\raggedright}p{1.8cm}
           >{\raggedright}p{1.8cm}
           >{\raggedright}p{1.9cm}
           l}
                & Post-BDT & Pair cand. & Pair cand. \\
               & singles rate & post-merging & in ROI \\
                & [\SI{}{\hertz}] & [\SI{}{\evts\per\year}] &  [\SI{}{\evts\per\year}] \\
           \hline  \noalign{\smallskip}
                PMT \ce{^{214}Bi}             & \SI{4.24}{}    & \num{6.80e4} & 24.0 \\
                PMT \ce{^{208}Tl}             & \SI{3.10}{}    & \num{4.68e4} & 13.6 \\
                PMT \ce{^{40}K}               & \SI{1.47}{}    & \num{2.41e4} & 11.6 \\
                Water \ce{^{214}Bi}           & \SI{8.18e-1}{} & \num{1.21e4} & 32.9 \\
                Water \ce{^{208}Tl}           & \SI{1.59e-2}{} & \num{0.21e4} & 0.78 \\
                
                \hline\noalign{\smallskip}            
                
                Cosmic \ce{^17N}  & \SI{5.78e-6}{} & 182  & 7.10\\ 
                Atm. NCQE     & \SI{1.13e-6}{} & 35.6 & 10.7 \\ 
                Cosmic n          & \SI{4.48e-4}{} & 1411 & 69.6\\
                PMTs$(\alpha,n)$      & \SI{1.51e-4}{} & 786 & 0.35 \\
                $\ce{^18O}(\alpha,n)$ & \SI{4.58e-7}{} & 14.5 & 12.8\\
            \hline\noalign{\smallskip}
            \hline\noalign{\smallskip}
                Reactor                 & \SI{7.61e-6}{} & 240 & 166 \\
                Geoneutrinos       & \SI{9.36e-6}{} & 295 & 221 \\
            \end{tabular}
        \end{table}
      
        Using all three cuts $\dt$, $\dr$ and $\tau_{\text{BDT}}$, a suppression factor on the order of $10^{-5}$ is observed for accidental backgrounds, whereas 54\% of geo- and 63\% of reactor antineutrinos are preserved. 
        Additional studies based on a simpler analysis show that a similar order of magnitude in suppression factors can be achieved with an optimized two-dimensional cut on $d_\text{wall}$ and $\chi^2$, supporting the conclusions from this more sophisticated study.
        
    \subsection{Box analysis}
    \label{subs:box}
       \begin{table*}[!htb]
        \centering
        \caption{Results of ROI optimization in the phase space of selection cuts necessary for antineutrino candidates selection.}
        \label{tab:optimization}
        \begin{tabular}{llllllllll}
            \hline\noalign{\smallskip}
            Signal & $\Ep$ & $\Ed$ & $z$ & $\rho$ & $\dt$ & $\dr$ & S & B  & ${S}/{\sqrt{S+B}}$ \\
            definition & range [n$_{400}$] & range [n$_{400}$] & [\SI{}{\meter}] & [\SI{}{\meter}] & [\SI{}{\micro\second}] & [\SI{}{\meter}] & [\SI{}{\evts\per\year}] & [\SI{}{\evts\per\year}] & \\
            \noalign{\smallskip}\hline\noalign{\smallskip}
            Geoneutrinos    & (200, 550)  & (400,600) & 31 & 7 & 700 & 1.4 & 221 & 169 & 11.2\\
            Reactor         & (500, 1200) & (400,600) & 31 & 7 & 700 & 1.4 & 124 & 189 & 7.02\\
            Geo + reactor   & (200, 1200) & (400,600) & 31 & 7 & 700 & 1.4 & 387 & 280 & 15.0\\
            \noalign{\smallskip}\hline
        \end{tabular}
    \end{table*}
    We perform a box optimization of selection cuts in eight dimensions: prompt and delayed energy thresholds and upper limits (expressed in $n_{400}$), $\rho$ ($\perp$ cylinder axis) and $z$ ($\parallel$ cylinder axis) to define a fiducial volume in 2-D, and $\dt$ and $\dr$. 
    We optimize a ROI to maximize the signal-to-background ratio, specifically $S/\sqrt{S+B}$ for three cases:  
    \begin{enumerate}
        \item when geoneutrinos are considered signal, and everything else including reactor antineutrinos are background;
        \item when reactor antineutrinos are considered signal, and everything else including geoneutrinos are counted towards background; 
        \item both geo- and reactor neutrinos are signal, and the rest is background. 
    \end{enumerate}
    The results of this optimization are summarized in Table~\ref{tab:optimization}. 
    A high signal-to-background ratio is achieved with the box analysis within one year for all three studied scenarios.
    The selection cuts in $\dt$, $\dr$, delayed energy range, and fiducial volume have converged to the same values for all three scenarios, and only the prompt energy range is dependent on the target signal.
     Comparing the values in the third and fourth columns in Table~\ref{tab:eff_rates} shows the impact of the optimized selection cuts on the number of pair candidates. Dedicated studies have been performed to study the impact of applying the BDT cut to the spatial and energy distributions of the events in the ROI. No significant change in the spectral shapes has been observed with varying degrees of BDT threshold, allowing us to use pre-BDT energy spectra as the PDFs for the spectral likelihood fit.
  
   Using the box analysis, we obtain a signal-to-background ratio $S/\sqrt{S+B}$ of 7.02 for reactor signal, 11.2 for geoneutrinos, and 15.0 for the summed antineutrino signal after one year of data taking. These significances do not include any systematic uncertainties on the individual contributions. 
   
    \subsection{Sensitivity analysis} 
        \label{subs:sensi}
    
\begin{figure}[!htb]
            \resizebox{0.49\textwidth}{!}{
                \includegraphics{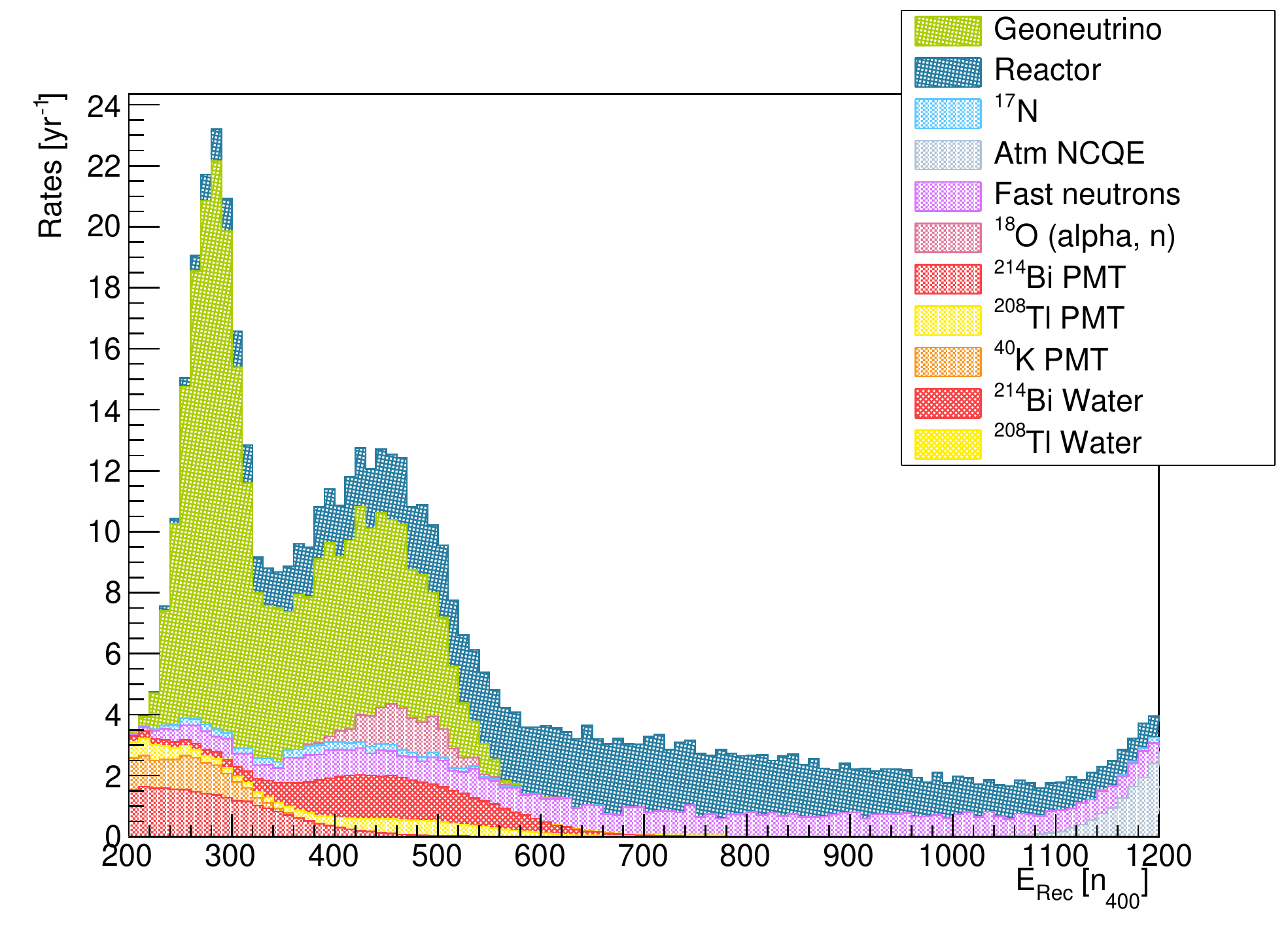}    
            }
            \caption{
                Stacked PDF spectra of all signal and background contributions in the ROI optimized for geo and reactor combined signal.
            }
            \label{fig:stacked_PDFs}
        \end{figure}

        The box analysis allows us to extract the rates $\lambda_i$ of each signal and background inside the optimized cuts. 
        We then create a toy experiment by randomly sampling events from PDF of each spectrum, assuming Poisson fluctuations for all the rates, and perform the likelihood fit.
        Figure~\ref{fig:stacked_PDFs} depicts the stacked spectra of all the PDFs scaled in correspondence with the expected rates in the ROI optimized for geo and reactor combined signal. 
        In the following section, the sensitivity of two analyses is presented: first, the simultaneous extraction of the geo- and reactor signals; second, the simultaneous extraction of the \ce{Th} and \ce{U} contributions to the geoneutrino signal. 
        A 2-D spectral fit is performed using a grid search on target signals $(S_\text{geo}, S_\text{rea})$ or $(S_\text{Th}, S_\text{U})$.
        For each couple of $S_i$ tested, a binned histogram is created by scaling each component of the PDF spectra. 
        Then a negative log-likelihood test against the toy experiment is performed while marginalizing on each background component $\lambda_B$.  
        The marginalization is computed between $\lambda_B\pm10\sqrt{\lambda_B}$ (or $[0, \lambda_B+10\sqrt{\lambda_B}]$ to ensure that all rates are positives), using the extended Newton-Cotes formulas, or trapezoidal rule\,\cite{press2002numerical}. 
        We extract an average negative log-likelihood for each couple of $S_i$ tested distributed from a $\chi^2$ law with two degrees of freedom from which we can extract the sensitivity of each analysis. 
        The minimum value corresponds to our best fit.
        To evaluate the sensitivity of this procedure, \num{1000} toy experiments are generated, each corresponding to one year of data taking.

\section{Results} 
\label{sec:results}

    \subsection{\Theia{}-25 one-year signal sensitivity}
    \label{subs:signal_sensitivity}
    
\begin{figure}[!htb]
            \resizebox{0.49\textwidth}{!}{
                \includegraphics{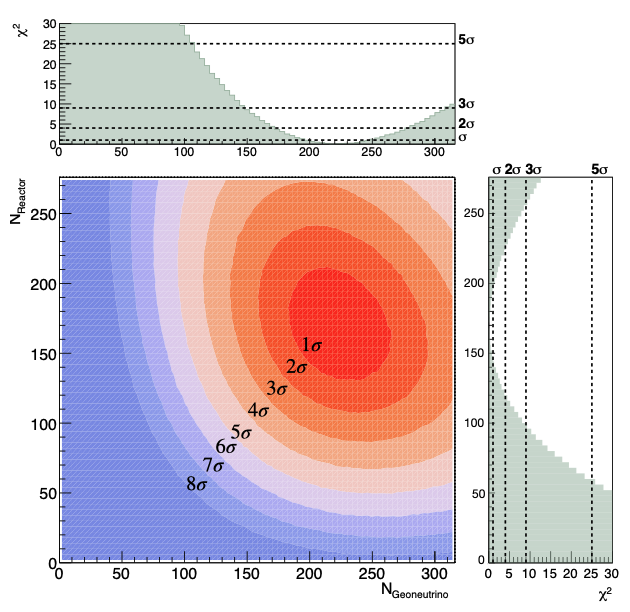}
            }
            \caption{
                2-D confidence level of the geoneutrino versus reactor antineutrinos flux after one year of data taking, estimated with \num{1000} toy experiments.
                The confidence level are drawn for $[1\sigma, 8\sigma]$.
                The individual projection for each contribution at the best fit position is also shown with the statistical significance as dashed line.
            }
            \label{fig:SignificanceNGeo}
        \end{figure}

\begin{figure}[!htb]
            \resizebox{0.49\textwidth}{!}{
                \includegraphics{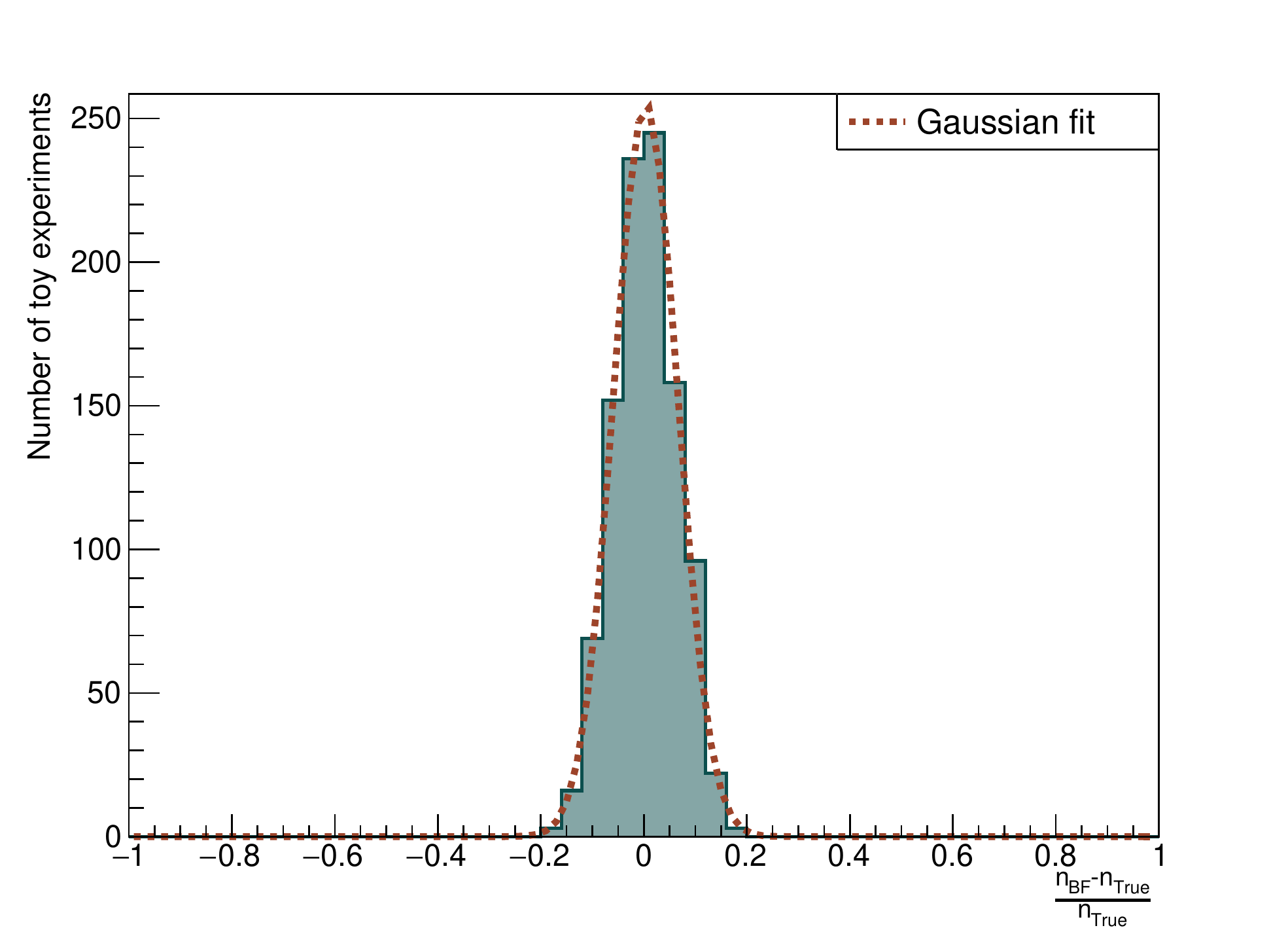}
            }
            \caption{The relative difference between the best fit value and the true value of the number of geoneutrinos, obtained with \num{1000} toy experiments.}
            \label{fig:Flux_precision}
        \end{figure}

        In this section, we present the results of two analyses: 
        i) extraction of the number of geo- and reactor antineutrinos, with geoneutrino energy spectrum based on the fixed \ce{U}/\ce{Th} ratio, 
        ii) extraction of the number of individual contributions of U and Th geoneutrinos, with two separate energy spectra for \ce{U} and \ce{Th}. 
        Figure~\ref{fig:SignificanceNGeo} shows the simultaneous extraction of the number of geo- and reactor antineutrinos with fixed \ce{U}/\ce{Th} ratio.
        The best fit values of 1000 toy experiments are $218^{+28}_{-20}$ and $170^{+24}_{-20}$ for geo- and reactor antineutrinos, respectively. This corresponds to 40\% and 44\% selection efficiencies for geo- and reactor antineutrinos.
        The contours in Fig.~\ref{fig:SignificanceNGeo} correspond to the $[1\sigma, 8\sigma]$ confidence levels. 
        The no-signal hypothesis for both geo- and reactor neutrinos is rejected at more than $8\sigma$ in a year.
        The precision of the fitting procedure has been evaluated using the relative difference between the best fit value and the true value.
        Figure~\ref{fig:Flux_precision} shows this difference for each toy experiment in the space of geoneutrinos. 
        A Gaussian fit is applied to check that the fitting procedure does not introduce any systematic shifts (less than 0.004) and to extract the standard deviation $\sigma_\text{flux}$ quantifying the uncertainty on the antineutrino interactions in a one year experiment. 
        The standard deviation values from the fits are $\sigma_\text{flux}^\text{geo} = 6.72\%$ for geoneutrinos and $\sigma_\text{flux}^\text{rea} = 8.55\%$ for reactors. 
        
\begin{figure}[!htb]
            \resizebox{0.49\textwidth}{!}{
                \includegraphics{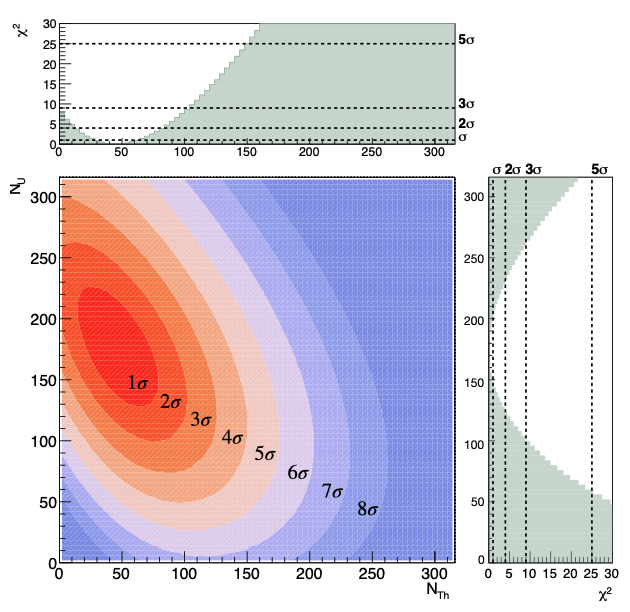}
            }
            \caption{
                2-D confidence level of the number of \ce{Th} versus \ce{U} antineutrinos detected after one year of data taking, estimated with \num{1000} toy experiments. 
                The confidence level are drawn for $[1\sigma, 8\sigma]$.
                The individual projections for each contribution at the best-fit value are also shown with the statistical significance as dashed lines.                    }
            \label{fig:exFitThU}
        \end{figure}

        Using the selection cuts optimized for the geoneutrino signal only (see Table~\ref{tab:optimization}), we performed a spectral fit to extract individual contributions of \ce{U} and \ce{Th} geoneutrino fluxes. 
        Marginalizing on all backgrounds, including reactor antineutrinos, we extract the best fit and confidence level contours between $[1\sigma, 8\sigma]$, as represented in Fig.~\ref{fig:exFitThU}.
        The best fit values for the spectral fit of individual contributions of \ce{Th} and \ce{U} yields $N_\text{Th}=39^{+18}_{-15}$ and $N_\text{U}=180^{+26}_{-22}$.
        $N_{\ce{U}}$ null hypothesis can be rejected with more than 5$\sigma$ confidence level within one year, while $N_{\ce{Th}}$ null hypothesis can be rejected at more than 2$\sigma$.
        Table~\ref{tab:measuredNIU} provides the summary of measured signal events, along with conversion to NIU using the efficiency of the selection cuts in total geoneutrinos (40\%) or from the individual \ce{U} (43\%) and \ce{Th} (31\%). The input model values are provided as well for the reference.

          \begin{table}[!htb]
            \centering
            \begin{tabular}{r|c|c|c}
                 & $N_\text{meas}$[\SI{}{\evts/\year}]  & $S_\text{meas}$ [NIU] & $S_\text{input}$ [NIU] \\
            \hline\noalign{\smallskip}
                \ce{U}   & $180^{+26}_{-22}$  & $35.5^{+5.1}_{-4.3}$ & 35.8  \\
                \ce{Th}  & $39^{+18}_{-15}$   & $10.6^{+4.9}_{-4.1}$ & 10.5   \\
                \ce{Geo} & $218^{+28}_{-20}$  & $45.9^{+5.9}_{-4.2}$ & 46.3  \\
            \hline\noalign{\smallskip}
            \end{tabular}
            \caption{The results of the sensitivity studies performed with likelihood fit with free U/Th ratio, for  U and Th values, and with fixed U/Th ratio for geoneutrinos. The first column presents best fit values in raw number of events per year, the second column is the result of conversion of the results into NIU that can be compared to other experiments, and finally third column is the input model that was used in the simulation for an easy reference.}
            \label{tab:measuredNIU}
        \end{table}
        
    \subsection{Geophysics interpretation of the sensitivity}
    

        The individual number of \ce{U} and \ce{Th} geoneutrinos can be used to evaluate the sensitivity of geophysical parameters at SURF.
        But combining both \ce{U} and \ce{Th} measurements is non-trivial as the $1\sigma$ error intervals are asymmetrical and highly correlated, as seen on the contours in Fig.\,\ref{fig:exFitThU}. 
        However, these contours are close to elliptical and equally spaced, meaning the correlated errors can be approximated as a bivariate Gaussian. 
        Since the best fit values lie only slightly skewed from the center of these ellipsis, in the following section the errors are treated as symmetrical, using the best fit values reported in Table~\ref{tab:measuredNIU}, but choosing the largest errors in order to be conservative.
        Therefore, the error matrix $\boldsymbol{\sigma^2}$ can be approximated by a bivariate Gaussian in the form:
        \begin{equation}
            \boldsymbol{\sigma^2} = 
            \begin{pmatrix}
                \sigma^2_\text{Th} & \rho\sigma_\text{Th}\sigma_\text{U} \\
                \rho\sigma_\text{Th}\sigma_\text{U} & \sigma^2_\text{U}
            \end{pmatrix},
        \end{equation}
        where $\sigma_\text{Th}=18$, $\sigma_\text{U}=26$, and $\rho=-0.75$ extracted from Fig.\,\ref{fig:exFitThU}.
        Using this parametrization, the total geoneutrino signal at SURF is $S_\text{geo} = 46.1\pm3.6$\,NIU.
        
        The $N_{\ce{Th}}/N_{\ce{U}}$ ratio is proportional to (\ce{Th}/\ce{U}) mass ratio\,\cite{PhysRevD.82.093006}, and hence is an important parameter for comparison with geological models. 
        For a one year exposure in \Theia{}-25, we obtain  $N_{\ce{Th}}/N_{\ce{U}} = 0.22\pm0.13$, which corresponds to \ce{Th}/\ce{U} mass ratio of $4.3\pm2.6$. 
        For comparison, the only preliminary measurement of (\ce{Th}/\ce{U}) mass ratio with geoneutrinos is $4.1^{+5.5}_{-2.0}$, performed by the KamLAND collaboration\,\cite{KamLAND_preliminary}. 
        From our sensitivity studies, we expect to discard the null hypothesis of (\ce{Th}/\ce{U}) mass ratio with $3\sigma$ confidence level with three years of data, and the uncertainty of this measurement to be reduced to 15\% with ten years of data taking. 

        Finally, the mantle contribution can be extracted by subtracting the predictions of the crust signal. 
        It is expected that by the time of \Theia{}-25 data taking, a detailed survey of the crust surrounding SURF would provide the more refined model of local crust and hence lower the uncertainties on the total crust predictions, as it has been done for Borexino and KamLAND\,\cite{PhysRevD.86.033004}. 
        Assuming a presumably positive correlation in the crust error matrix $\boldsymbol{\sigma}_c$, the mantle error matrix would be augmented such as $\boldsymbol{\sigma}^2_m = \boldsymbol{\sigma}^2+\boldsymbol{\sigma}^2_c$.
        Therefore using uncertainties on the crust fluxes, similar to KamLAND, of 8.3\% on \ce{Th} and 7.0\% on \ce{U}, the estimated mantle signal would be $S_\text{mantle} = 9.0\pm[4.2, 4.5]$\,NIU depending on $\rho_c\in[0, 1]$. 




  \subsection{Systematic uncertainties}      
        \Theia{}-25 one-year sensitivity has also been evaluated using the systematic uncertainties on the backgrounds, described in Table~\ref{tab:systs}. 
        A conservative 10\% uncertainty is assumed on the accidental background from the contaminants inside the PMTs' glass and WbLS, 
        which also includes the contribution from the unknown radiogenic activity from the cavern rocks. 
        Nearby survey of \Theia{}-25 potential location at SURF for the LUX experiment\,\cite{Akerib:2014rda} implies possible increase on the singles rate of $\beta$ and $\gamma$-rays from 0.1 to 10\,\SI{}{\kilo\hertz}.
        A conservative 25\% uncertainty on the $(\alpha, n)$\,\cite{PhysRevC.7.1356} cross-section is also used for neutron production inside WbLS and PMTs' glass.
        The uncertainty on the cosmogenic background from either \ce{^17N} activation and fast neutrons from invisible muons comes from the measured muon flux at SURF, quoted at 3\%\,\cite{Muon_flux_Majorana}.
        However, the spallation production model for \ce{^17N} and fast neutrons contains significant uncertainties, and a conservative 100\% has been taken. 
        This large variation also includes the unknown contribution from radiogenic neutrons from the cavern rocks.
        The atmospheric neutrino flux uncertainty is taken as 100\%.
        In addition, shape uncertainties have also been considered when throwing toy experiments. 
        Values quoted for the shape uncertainties are conservative and similar to concurrent sensitivity studies\,\cite{Han:2015roa}.
        \begin{table}[!htb]
            \centering
            \begin{tabular}{r|c|c}
                & Norm & Shape \\
                \cline{2-3}
                Accidental & 10\% & - \\
                \hline
                Cosmic \ce{^17N} & 100\% & 20\% \\
                Cosmic n & 100\% & 20\% \\
                \hline 
                $\ce{^18O}(\alpha, n)$ & 25\% & 25\% \\
                PMTs$(\alpha, n)$ & 25\% & 25\% \\
                \hline 
                Atm. NCQE & 100\% & 100\% \\
            \end{tabular}
            \caption{Summary of systematic uncertainties considered for \Theia{}-25 sensitivity to geo- and reactor antineutrinos.}
            \label{tab:systs}
        \end{table}
        We build another set of toy experiments, where background contributions are varied over the extended range to include systematic uncertainties. 
        Likelihood fit analysis, including all the assumptions are unchanged from the above analysis. 
        The best fit values for the number of geo- and reactor antineutrinos are $220^{+30}_{-24}$ and $168^{+26}_{-24}$ respectively. 
        The no-signal hypothesis for detecting the geo- and reactor antineutrinos flux simultaneously after one year of data taking is still rejected at the $8\sigma$ level. 
        This confirms that our analysis to extract number of geo- and reactor neutrinos is robust in respect to the background assumptions. 
        Moreover, we repeat our sensitivity analysis with an additional set of toy experiments, where we vary the contributions of geo- and rector antineutrinos beyond the Poisson fluctuations for the expected rates. 
        For reactor signal, we add 2.7\% to account for uncertainties on the average survival probability of electron antineutrinos, $\langle P_\text{ee} \rangle = 1 - 1/2\left(\cos^4\theta_{13}\sin^2(2\theta_{12})+\sin^2(2\theta_{13})\right)$, combined with the fission isotope emission spectra and the core composition of nuclear reactor. 
        For geoneutrinos, we consider as much as 23.8\% additional variation\,\cite{Dye:2015bsw}, when building the toy experiments to account for uncertainties of oscillation parameters and geological models. 
        
        Following the same procedure, as described above, we obtain the relative difference between the best fit value and the true value.
        The standard deviation values from the gaussian fits are $\sigma_\text{flux}^\text{geo} = 8.7\%$ for geoneutrinos and $\sigma_\text{flux}^\text{rea} = 10.1\%$ for reactors, showing only slight reduction in the expected sensitivity. 
        Finally, we apply the systematic uncertainties in the same way, when building additional set of toy experiments for \ce{Th} vs \ce{U} antineutrinos likelihood fit. 
        The best fit values for the spectral fit of individual contributions of \ce{Th} and \ce{U} yields $N_\text{Th}=40^{+26}_{-22}$ and $N_\text{U}=180^{+30}_{-24}$.
        Assuming previous symmetric error matrix, but now $\sigma_\text{Th}=26$, $\sigma_\text{U}=30$, and $\rho=-0.75$, the total geoneutrino flux at SURF is $S_\text{geo} = 46.4\pm4.7$\,NIU.
        The measured $N_{\ce{Th}}/N_{\ce{U}}=0.22 \pm 0.17$ corresponds to \ce{Th}/\ce{U} mass ratio of $4.3\pm 2.6$. 
        Finally, the mantle signal extracted from individual \ce{U} and \ce {Th} contributions fit is $S_\text{mantle} = 9.3\pm[5.2, 5.4]$\,NIU depending on $\rho_c\in[0, 1]$.

\section{Conclusions}

Up to date, only two experiments in the world, Borexino and KamLAND, have observed a combined total of 216.6 geoneutrinos in the span of 12 years and 14 years, correspondingly. 
As presented above, \Theia{}-25 will coincidentally observe the same number of geoneutrinos within just one year of data-taking. 
For this analysis, we considered one of the possible designs for the \Theia{} detector, a 25-ktonne detector filled with WbLS placed at SURF.  
\Theia{}-25 will provide the first high-statistics measurement of geoneutrinos in North America: $218^{+28}_{-20}$ events per year. 
Due to variations of the crust thickness, the geoneutrino flux measurements at different geographical locations will help separate a much less position-dependent mantle contribution \cite{SMIRNOV2019103712}. 
The geoneutrino measurement in \Theia{}-25 after one year of data taking assuming $(\ce{Th}/\ce{U})=4.24$ corresponds to $S_\text{geo} = 45.9^{+5.9}_{-4.2}$\,NIU. 
We also demonstrate the sensitivity towards fitting individual Th and U contributions, with best fit values of $N_\text{Th}=39^{+18}_{-15}$ and $N_\text{U}=180^{+26}_{-22}$. 
We obtain $(\ce{Th}/\ce{U})=4.3\pm2.6$ after one year of data taking, and within ten years, the relative precision of the $(\ce{Th}/\ce{U})$ mass ratio will be reduced to 15\%. 
A global assessment of the (\ce{Th}/\ce{U}) mass ratio of the primitive mantle could give insight into the Earth's early evolution and its differentiation. 
(\ce{Th}/\ce{U}) concentration in the outermost Earth's crust can be sampled directly, but determining these concentrations in the mantle can be done only with the geoneutrino flux measurements. 
We evaluate the mantle signal at \Theia{}-25 to be $S_\text{mantle} = 9.0\pm[4.2, 4.5]$\,NIU from the fit results of individual Th and U contributions and depending on the correlation between the crust theoretical prediction.
Assuming homogeneous radioactive element concentrations in the mantle, \ce{Th}/\ce{U}=3.9, and \ce{K}/\ce{U}=$10^4$, our mantle measurement corresponds to $H=10.7\pm[5.0, 5.3]$\,\SI{}{\tera\watt} radiogenic heat.

Moreover, while we discuss sensitivity in \Theia{}-25 in this paper, in further studies, we plan to explore the potential improvements in sensitivity with higher antineutrino statistics achievable in \Theia{}-100, as well as with extracting the directional information of IBD interactions and applying it to the antineutrino search. 
The two existing measurements confirm the general validity of different geological models predicting the Th and U abundances in the Earth. With \Theia{} and other large experiments providing new precise measurements, we will enter a new era of geoneutrino measurements informing geoscience. 

\section{Acknowledgements}

This work was performed under the auspices of the U.S. Department of Energy by Lawrence Berkeley National Laboratory under Contract DE-AC02-05CH11231. The project was funded by the U.S. Department of Energy, National Nuclear Security Administration, Office of Defense Nuclear Nonproliferation Research and Development (DNN R\&D). We also thank Tanner Kaptanoglu for detailed review of the manuscript and insightful suggestions.

\printbibliography 

@article{Li:2014sea,
    author = "Li, Shirley Weishi and Beacom, John F.",
    title = "{First calculation of cosmic-ray muon spallation backgrounds for MeV astrophysical neutrino signals in Super-Kamiokande}",
    eprint = "1402.4687",
    archivePrefix = "arXiv",
    primaryClass = "hep-ph",
    doi = "10.1103/PhysRevC.89.045801",
    journal = "Phys. Rev. C",
    volume = "89",
    pages = "045801",
    year = "2014"
}

@article{Akerib:2014rda,
    author = "Akerib, D. S. and others",
    title = "{Radiogenic and Muon-Induced Backgrounds in the LUX Dark Matter Detector}",
    eprint = "1403.1299",
    archivePrefix = "arXiv",
    primaryClass = "astro-ph.IM",
    doi = "10.1016/j.astropartphys.2014.07.009",
    journal = "Astropart. Phys.",
    volume = "62",
    pages = "33--46",
    year = "2015"
}

@article{Han:2015roa,
    author = "Han, Ran and Li, Yu-Feng and Zhan, Liang and McDonough, William F. and Cao, Jun and Ludhova, Livia",
    title = "{Potential of Geo-neutrino Measurements at JUNO}",
    eprint = "1510.01523",
    archivePrefix = "arXiv",
    primaryClass = "physics.ins-det",
    doi = "10.1088/1674-1137/40/3/033003",
    journal = "Chin. Phys. C",
    volume = "40",
    number = "3",
    pages = "033003",
    year = "2016"
}

@article{PhysRevD.86.033004,
  title = {Mantle geoneutrinos in KamLAND and Borexino},
  author = {Fiorentini, G. and Fogli, G. L. and Lisi, E. and Mantovani, F. and Rotunno, A. M.},
  journal = {Phys. Rev. D},
  volume = {86},
  issue = {3},
  pages = {033004},
  numpages = {11},
  year = {2012},
  month = {Aug},
  publisher = {American Physical Society},
  doi = {10.1103/PhysRevD.86.033004},
  url = {https://link.aps.org/doi/10.1103/PhysRevD.86.033004}
}

@article{PhysRevC.7.1356,
  title = {Total Neutron Yield from the Reactions $^{13}\mathrm{C}(\ensuremath{\alpha}, n)^{16}\mathrm{O}$ and $^{17,18}\mathrm{O}(\ensuremath{\alpha}, n)^{20,21}\mathrm{Ne}$},
  author = {Bair, J. K. and Haas, F. X.},
  journal = {Phys. Rev. C},
  volume = {7},
  issue = {4},
  pages = {1356--1364},
  numpages = {0},
  year = {1973},
  month = {Apr},
  publisher = {American Physical Society},
  doi = {10.1103/PhysRevC.7.1356},
  url = {https://link.aps.org/doi/10.1103/PhysRevC.7.1356}
}

@article{theia_wp,
    author = {Askins, M. and others},
    year = {2020},
    month = {05},
    pages = {},
    title = {Theia: an advanced optical neutrino detector},
    volume = {80},
    journal = {Eur. Phys. J. C},
    doi = {10.1140/epjc/s10052-020-7977-8}
}

@article{wbls,
	author         = "Yeh, M. and Hans, S. and Beriguete, W. and Rosero, R. and Hu, L. and Hahn, R. L. and Diwan, M. V. and Jaffe, D. E. and Kettell, S. H. and Littenberg, L.",
	title          = "{A new water-based liquid scintillator and potential applications}",
	journal        = "Nucl. Instrum. Meth. A",
	volume         = "660",
	year           = "2011",
	pages          = "51-56",
	doi            = "10.1016/j.nima.2011.08.040"
}

@article{snoplus,
	author         = "Andringa, S. and others",
	title          = "{Current Status and Future Prospects of the SNO+ Experiment}",
	collaboration  = "SNO+",
	journal        = "Adv. High Energy Phys.",
	volume         = "2016",
	year           = "2016",
	pages          = "6194250",
	doi            = "10.1155/2016/6194250",
}

@Article{drew_wbls,
    author ="Onken, Drew R. and Moretti, Federico and Caravaca, Javier and Yeh, Minfang and Orebi Gann, Gabriel D. and Bourret, Edith D.",
    title  ="Time response of water-based liquid scintillator from X-ray excitation",
    journal  ="Mater. Adv.",
    year  ="2020",
    volume  ="1",
    issue  ="1",
    pages  ="71-76",
    publisher  ="RSC",
    doi  ="10.1039/D0MA00055H"
}

@Article{chess_wbls,
   author = "Caravaca, J. and Land, B.J. and Yeh, M. and Orebi Gann, G.D.",
    title = "{Characterization of water-based liquid scintillator for Cherenkov and scintillation separation}",
    journal  ="Eur. Phys. J. C ",
    year  ="2020",
    volume  ="80",
    pages  ="867",
    doi  ="https://doi.org/10.1140/epjc/s10052-020-8418-4"
}

@article{juno_2022,
title = {JUNO physics and detector},
journal = {Progress in Particle and Nuclear Physics},
volume = {123},
pages = {103927},
year = {2022},
issn = {0146-6410},
doi = {https://doi.org/10.1016/j.ppnp.2021.103927},
url = {https://www.sciencedirect.com/science/article/pii/S0146641021000880},
}

@article{geant4,
	author         = "Agostinelli, S. and others",
	title          = "{GEANT4: A Simulation toolkit}",
	collaboration  = "GEANT4",
	journal        = "Nucl. Instrum. Meth. A",
	volume         = "506",
	year           = "2003",
	pages          = "250-303",
	doi            = "10.1016/S0168-9002(03)01368-8",
	reportNumber   = "SLAC-PUB-9350, FERMILAB-PUB-03-339"
}

@misc{ratpac,
%	title = "{RAT-PAC User's Guide}",
	howpublished = {RAT-PAC User's Guide\url{https://rat.readthedocs.io/en/latest/}}
}

@misc{NeuCBOT,
%	title = "{NeuCBOT v-2.0}",
	howpublished = {NeuCBOT v-2.0 \url{https://github.com/shawest/neucbot}}
}

@article{PhysRevD.91.065002,
  title = {Reference worldwide model for antineutrinos from reactors},
  author = {Baldoncini, Marica and Callegari, Ivan and Fiorentini, Giovanni and Mantovani, Fabio and Ricci, Barbara and Strati, Virginia and Xhixha, Gerti},
  journal = {Phys. Rev. D},
  volume = {91},
  issue = {6},
  pages = {065002},
  numpages = {16},
  year = {2015},
  publisher = {American Physical Society},
  doi = {10.1103/PhysRevD.91.065002},
  url = {https://link.aps.org/doi/10.1103/PhysRevD.91.065002}
}

@misc{nobel_2015,
%   howpublished = {https://www.nobelprize.org/uploads/2017/09/advanced-physicsprize2015.pdf},
%author = {Class for Physics of the Royal Swedish Academy of Sciences},
%journal = {Royal Swedish Academy of Sciences},
howpublished = {Royal Swedish Academy of Sciences - Scientific Background on the Nobel Prize in Physics 2015-
Neutrino Oscillations},
}

@article{Bilenky:1978nj,
    author = "Bilenky, Samoil M. and Pontecorvo, B.",
    title = "{Lepton Mixing and Neutrino Oscillations}",
    doi = "10.1016/0370-1573(78)90095-9",
    journal = "Phys. Rept.",
    volume = "41",
    pages = "225--261",
    year = "1978"
}

@article{PROSPECT:2020sxr,
    author = "Andriamirado, M. and others",
    collaboration = "PROSPECT",
    title = "{Improved short-baseline neutrino oscillation search and energy spectrum measurement with the PROSPECT experiment at HFIR}",
    eprint = "2006.11210",
    archivePrefix = "arXiv",
    primaryClass = "hep-ex",
    doi = "10.1103/PhysRevD.103.032001",
    journal = "Phys. Rev. D",
    volume = "103",
    number = "3",
    pages = "032001",
    year = "2021"
}

@article{STEREO:2019ztb,
    author = "Almaz\'an, H. and others",
    collaboration = "STEREO",
    title = "{Improved sterile neutrino constraints from the STEREO experiment with 179 days of reactor-on data}",
    eprint = "1912.06582",
    archivePrefix = "arXiv",
    primaryClass = "hep-ex",
    doi = "10.1103/PhysRevD.102.052002",
    journal = "Phys. Rev. D",
    volume = "102",
    number = "5",
    pages = "052002",
    year = "2020"
}

@misc{ferrara_reactors,
	title = "{A reference worldwide model for antineutrinos from reactors}",
	howpublished = {A reference worldwide model for antineutrinos from reactors \url{https://www.fe.infn.it/antineutrino/}}
}

@misc{snoplus_private,
	author = "{The SNO+ collaboration}",
	howpublished = "Private communication"
}

@misc{watchman_private,
	author = "{The WATCHMAN collaboration}",
	howpublished = "Private communication"
}

@misc{bnl_private,
	howpublished = "Private communication with Minfang Yeh (Brookhaven National Laboratory)"
}

@article{Cleveland:1998nv,
    author = "Cleveland, B. T. and Daily, Timothy and Davis, Jr., Raymond and Distel, James R. and Lande, Kenneth and Lee, C. K. and Wildenhain, Paul S. and Ullman, Jack",
    title = "{Measurement of the solar electron neutrino flux with the Homestake chlorine detector}",
    doi = "10.1086/305343",
    journal = "Astrophys. J.",
    volume = "496",
    pages = "505--526",
    year = "1998"
}

@article{Reines:1965qk,
    author = "Reines, F. and Crouch, M. F. and Jenkins, T. L. and Kropp, W. R. and Gurr, H. S. and Smith, G. R. and Sellschop, J. P. F. and Meyer, B.",
    title = "{Evidence for high-energy cosmic ray neutrino interactions}",
    doi = "10.1103/PhysRevLett.15.429",
    journal = "Phys. Rev. Lett.",
    volume = "15",
    pages = "429--433",
    year = "1965"
}

@article{Bionta:1987qt,
    author = "Bionta, R. M. and others",
    title = "{Observation of a Neutrino Burst in Coincidence with Supernova SN 1987a in the Large Magellanic Cloud}",
    reportNumber = "UCI-NEUTRINO-87-10",
    doi = "10.1103/PhysRevLett.58.1494",
    journal = "Phys. Rev. Lett.",
    volume = "58",
    pages = "1494",
    year = "1987"
}

@article{Kamiokande-II:1987idp,
    author = "Hirata, K. and others",
    editor = "Wali, K. C.",
    collaboration = "Kamiokande-II",
    title = "{Observation of a Neutrino Burst from the Supernova SN 1987a}",
    reportNumber = "UT-ICEPP-87-01, UPR-142E",
    doi = "10.1103/PhysRevLett.58.1490",
    journal = "Phys. Rev. Lett.",
    volume = "58",
    pages = "1490--1493",
    year = "1987"
}

@article{Borexino:2013tnm,
    author = "Bellini, G. and others",
    collaboration = "Borexino",
    title = "{Measurement of geo-neutrinos from 1353 days of Borexino}",
    eprint = "1303.2571",
    archivePrefix = "arXiv",
    primaryClass = "hep-ex",
    doi = "10.1016/j.physletb.2013.04.030",
    journal = "Phys. Lett. B",
    volume = "722",
    pages = "295--300",
    year = "2013"
}

@article{Araki:2005qa,
    author = "Araki, T. and others",
    title = "{Experimental investigation of geologically produced antineutrinos with KamLAND}",
    doi = "10.1038/nature03980",
    journal = "Nature",
    volume = "436",
    pages = "499--503",
    year = "2005"
}

@article{K2K:2002icj,
    author = "Ahn, M. H. and others",
    collaboration = "K2K",
    title = "{Indications of neutrino oscillation in a 250 km long baseline experiment}",
    eprint = "hep-ex/0212007",
    archivePrefix = "arXiv",
    doi = "10.1103/PhysRevLett.90.041801",
    journal = "Phys. Rev. Lett.",
    volume = "90",
    pages = "041801",
    year = "2003"
}

@ARTICLE{hanohano,
       author = {{Learned}, John G. and {Dye}, Stephen T. and {Pakvasa}, Sandip},
        title = "{Hanohano: A Deep Ocean Anti-Neutrino Detector for Unique Neutrino Physics and Geophysics Studies}",
    %   journal = {arXiv e-prints},
     keywords = {High Energy Physics - Experiment, High Energy Physics - Phenomenology},
         year = 2008,
        month = oct,
          eid = {arXiv:0810.4975},
        % pages = {arXiv:0810.4975},
archivePrefix = {arXiv},
       eprint = {0810.4975},
 primaryClass = {hep-ex},
       adsurl = {https://ui.adsabs.harvard.edu/abs/2008arXiv0810.4975L},
      adsnote = {Provided by the SAO/NASA Astrophysics Data System}
}

@article{snowmass-OBD,
	author = {{Watanabe}, H. and {Inoue}, K. and {Sakai}, T. and {McDonough}, W. F. and {Abe}, N. and  {Araki}, E. and {Kasaya}, T. and {Kyo}, M. and {Sakurai}, N. and {Ueki}, K. and {Yoshida}, H.},
	title = {Snowmass2021-Letter of Interest Ocean Bottom Detector},
	journal = {Snowmass2021},
	year = {2021}
	}

@article{Shukla:2016nio,
    author = "Shukla, Prashant and Sankrith, Sundaresh",
    title = "{Energy and angular distributions of atmospheric muons at the Earth}",
    eprint = "1606.06907",
    archivePrefix = "arXiv",
    primaryClass = "hep-ph",
    doi = "10.1142/S0217751X18501750",
    journal = "Int. J. Mod. Phys. A",
    volume = "33",
    number = "30",
    pages = "1850175",
    year = "2018"
}

@article{Honda:2006qj,
    author = "Honda, Morihiro and Kajita, T. and Kasahara, K. and Midorikawa, S. and Sanuki, T.",
    title = "{Calculation of atmospheric neutrino flux using the interaction model calibrated with atmospheric muon data}",
    eprint = "astro-ph/0611418",
    archivePrefix = "arXiv",
    doi = "10.1103/PhysRevD.75.043006",
    journal = "Phys. Rev. D",
    volume = "75",
    pages = "043006",
    year = "2007"
}

@article{Super-Kamiokande:2019gzr,
    author = "Jiang, M. and others",
    collaboration = "Super-Kamiokande",
    title = "{Atmospheric Neutrino Oscillation Analysis with Improved Event Reconstruction in Super-Kamiokande IV}",
    eprint = "1901.03230",
    archivePrefix = "arXiv",
    primaryClass = "hep-ex",
    doi = "10.1093/ptep/ptz015",
    journal = "PTEP",
    volume = "2019",
    number = "5",
    pages = "053F01",
    year = "2019"
}

@book{press2002numerical,
  title={Numerical Recipes in C++: The Art of Scientific Computing},
  author={Press, W.H. and Press, W.H. and Teukolsky, S.A. and Vetterling, W.T. and Flannery, B.P.},
  isbn={9780521750332},
  lccn={2001052699},
  url={https://books.google.fr/books?id=mSLhDt\_XIUQC},
  year={2002},
  publisher={Cambridge University Press}
}

@article{Ankowski:2011ei,
    author = "Ankowski, Artur M. and Benhar, Omar and Mori, Takaaki and Yamaguchi, Ryuta and Sakuda, Makoto",
    title = "{Analysis of $\gamma$-ray production in neutral-current neutrino-oxygen interactions at energies above 200 MeV}",
    eprint = "1110.0679",
    archivePrefix = "arXiv",
    primaryClass = "nucl-th",
    doi = "10.1103/PhysRevLett.108.052505",
    journal = "Phys. Rev. Lett.",
    volume = "108",
    pages = "052505",
    year = "2012"
}

@article{PhysRevD.100.112009,
  title = {Measurement of neutrino and antineutrino neutral-current quasielasticlike interactions on oxygen by detecting nuclear deexcitation $\ensuremath{\gamma}$ rays},
  author = {Abe, K. and Akutsu, R. and Ali, A. and Alt, C. and Andreopoulos, C. and Anthony, L. and Antonova, M. and Aoki, S. and Ariga, A. and Ashida, Y. and Atkin, E. T. and Awataguchi, Y. and Ban, S. and Barbi, M. and Barker, G. J. and Barr, G. and Barry, C. and Batkiewicz-Kwasniak, M. and Beloshapkin, A. and Bench, F. and Berardi, V. and Berkman, S. and Berns, L. and Bhadra, S. and Bienstock, S. and Blondel, A. and Bolognesi, S. and Bourguille, B. and Boyd, S. B. and Brailsford, D. and Bravar, A. and Bravo Bergu\~no, D. and Bronner, C. and Bubak, A. and Buizza Avanzini, M. and Calcutt, J. and Campbell, T. and Cao, S. and Cartwright, S. L. and Catanesi, M. G. and Cervera, A. and Chappell, A. and Checchia, C. and Cherdack, D. and Chikuma, N. and Christodoulou, G. and Coleman, J. and Collazuol, G. and Cook, L. and Coplowe, D. and Cudd, A. and Dabrowska, A. and De Rosa, G. and Dealtry, T. and Denner, P. F. and Dennis, S. R. and Densham, C. and Di Lodovico, F. and Dokania, N. and Dolan, S. and Drapier, O. and Dumarchez, J. and Dunne, P. and Eklund, L. and Emery-Schrenk, S. and Ereditato, A. and Fernandez, P. and Feusels, T. and Finch, A. J. and Fiorentini, G. A. and Fiorillo, G. and Francois, C. and Friend, M. and Fujii, Y. and Fujita, R. and Fukuda, D. and Fukuda, R. and Fukuda, Y. and Gameil, K. and Giganti, C. and Golan, T. and Gonin, M. and Gorin, A. and Guigue, M. and Hadley, D. R. and Haigh, J. T. and Hamacher-Baumann, P. and Hartz, M. and Hasegawa, T. and Hastings, N. C. and Hayashino, T. and Hayato, Y. and Hiramoto, A. and Hogan, M. and Holeczek, J. and Hong Van, N. T. and Iacob, F. and Ichikawa, A. K. and Ikeda, M. and Ishida, T. and Ishii, T. and Ishitsuka, M. and Iwamoto, K. and Izmaylov, A. and Jamieson, B. and Jenkins, S. J. and Jes\'us-Valls, C. and Jiang, M. and Johnson, S. and Jonsson, P. and Jung, C. K. and Kabirnezhad, M. and Kaboth, A. C. and Kajita, T. and Kakuno, H. and Kameda, J. and Karlen, D. and Kasetti, S. P. and Kataoka, Y. and Katori, T. and Kato, Y. and Kearns, E. and Khabibullin, M. and Khotjantsev, A. and Kikawa, T. and Kim, H. and Kim, J. and King, S. and Kisiel, J. and Knight, A. and Knox, A. and Kobayashi, T. and Koch, L. and Koga, T. and Konaka, A. and Kormos, L. L. and Koshio, Y. and Kowalik, K. and Kubo, H. and Kudenko, Y. and Kukita, N. and Kuribayashi, S. and Kurjata, R. and Kutter, T. and Kuze, M. and Labarga, L. and Lagoda, J. and Lamoureux, M. and Laveder, M. and Lawe, M. and Licciardi, M. and Lindner, T. and Litchfield, R. P. and Liu, S. L. and Li, X. and Longhin, A. and Ludovici, L. and Lu, X. and Lux, T. and Machado, L. N. and Magaletti, L. and Mahn, K. and Malek, M. and Manly, S. and Maret, L. and Marino, A. D. and Martin, J. F. and Maruyama, T. and Matsubara, T. and Matsushita, K. and Matveev, V. and Mavrokoridis, K. and Mazzucato, E. and McCarthy, M. and McCauley, N. and McFarland, K. S. and McGrew, C. and Mefodiev, A. and Metelko, C. and Mezzetto, M. and Minamino, A. and Mineev, O. and Mine, S. and Miura, M. and Molina Bueno, L. and Moriyama, S. and Morrison, J. and Mueller, Th. A. and Munteanu, L. and Murphy, S. and Nagai, Y. and Nakadaira, T. and Nakahata, M. and Nakajima, Y. and Nakamura, A. and Nakamura, K. G. and Nakamura, K. and Nakayama, S. and Nakaya, T. and Nakayoshi, K. and Nantais, C. and Ngoc, T. V. and Niewczas, K. and Nishikawa, K. and Nishimura, Y. and Nonnenmacher, T. S. and Nova, F. and Novella, P. and Nowak, J. and Nugent, J. C. and O'Keeffe, H. M. and O'Sullivan, L. and Odagawa, T. and Okumura, K. and Okusawa, T. and Oser, S. M. and Owen, R. A. and Oyama, Y. and Palladino, V. and Palomino, J. L. and Paolone, V. and Parker, W. C. and Paudyal, P. and Pavin, M. and Payne, D. and Penn, G. C. and Pickering, L. and Pidcott, C. and Pinzon Guerra, E. S. and Pistillo, C. and Popov, B. and Porwit, K. and Posiadala-Zezula, M. and Pritchard, A. and Quilain, B. and Radermacher, T. and Radicioni, E. and Radics, B. and Ratoff, P. N. and Reinherz-Aronis, E. and Riccio, C. and Rondio, E. and Roth, S. and Rubbia, A. and Ruggeri, A. C. and Rychter, A. and Sakashita, K. and S\'anchez, F. and Schloesser, C. M. and Scholberg, K. and Schwehr, J. and Scott, M. and Seiya, Y. and Sekiguchi, T. and Sekiya, H. and Sgalaberna, D. and Shah, R. and Shaikhiev, A. and Shaker, F. and Shaykina, A. and Shiozawa, M. and Shorrock, W. and Shvartsman, A. and Smirnov, A. and Smy, M. and Sobczyk, J. T. and Sobel, H. and Soler, F. J. P. and Sonoda, Y. and Steinmann, J. and Suvorov, S. and Suzuki, A. and Suzuki, S. Y. and Suzuki, Y. and Sztuc, A. A. and Tada, M. and Tajima, M. and Takeda, A. and Takeuchi, Y. and Tanaka, H. K. and Tanaka, H. A. and Tanaka, S. and Thompson, L. F. and Toki, W. and Touramanis, C. and Tsui, K. M. and Tsukamoto, T. and Tzanov, M. and Uchida, Y. and Uno, W. and Vagins, M. and Valder, S. and Vallari, Z. and Vargas, D. and Vasseur, G. and Vilela, C. and Vinning, W. G. S. and Vladisavljevic, T. and Volkov, V. V. and Wachala, T. and Walker, J. and Walsh, J. G. and Wang, Y. and Wark, D. and Wascko, M. O. and Weber, A. and Wendell, R. and Wilking, M. J. and Wilkinson, C. and Wilson, J. R. and Wilson, R. J. and Wood, K. and Wret, C. and Yamada, Y. and Yamamoto, K. and Yanagisawa, C. and Yang, G. and Yano, T. and Yasutome, K. and Yen, S. and Yershov, N. and Yokoyama, M. and Yoshida, T. and Yu, M. and Zalewska, A. and Zalipska, J. and Zaremba, K. and Zarnecki, G. and Ziembicki, M. and Zimmerman, E. D. and Zito, M. and Zsoldos, S. and Zykova, A.},
  collaboration = {T2K Collaboration},
  journal = {Phys. Rev. D},
  volume = {100},
  issue = {11},
  pages = {112009},
  numpages = {19},
  year = {2019},
  month = {Dec},
  publisher = {American Physical Society},
  doi = {10.1103/PhysRevD.100.112009},
  url = {https://link.aps.org/doi/10.1103/PhysRevD.100.112009}
}

@article{Benhar:2005dj,
    author = "Benhar, Omar and Farina, Nicola and Nakamura, Hiroki and Sakuda, Makoto and Seki, Ryoichi",
    title = "{Electron- and neutrino-nucleus scattering in the impulse approximation regime}",
    eprint = "hep-ph/0506116",
    archivePrefix = "arXiv",
    doi = "10.1103/PhysRevD.72.053005",
    journal = "Phys. Rev. D",
    volume = "72",
    pages = "053005",
    year = "2005"
}

@article{Super-Kamiokande:2017yvm,
    author = "Abe, K. and others",
    collaboration = "Super-Kamiokande",
    title = "{Atmospheric neutrino oscillation analysis with external constraints in Super-Kamiokande I-IV}",
    eprint = "1710.09126",
    archivePrefix = "arXiv",
    primaryClass = "hep-ex",
    doi = "10.1103/PhysRevD.97.072001",
    journal = "Phys. Rev. D",
    volume = "97",
    number = "7",
    pages = "072001",
    year = "2018"
}

@article{Hayato:2009zz,
    author = "Hayato, Yoshinari",
   % editor = "Ankowski, Arthur and Sobczyk, Jan",
    title = "{A neutrino interaction simulation program library NEUT}",
    journal = "Acta Phys. Polon. B",
    volume = "40",
    pages = "2477--2489",
    year = "2009"
}

@article{Wolfenstein:1977ue,
    author = "Wolfenstein, L.",
    title = "{Neutrino Oscillations in Matter}",
    reportNumber = "COO-3066-102",
    doi = "10.1103/PhysRevD.17.2369",
    journal = "Phys. Rev. D",
    volume = "17",
    pages = "2369--2374",
    year = "1978"
}

@article{Strumia:2003zx,
    author = "Strumia, Alessandro and Vissani, Francesco",
    title = "{Precise quasielastic neutrino/nucleon cross-section}",
    eprint = "astro-ph/0302055",
    archivePrefix = "arXiv",
    reportNumber = "IFUP-TH-2003-2",
    doi = "10.1016/S0370-2693(03)00616-6",
    journal = "Phys. Lett. B",
    volume = "564",
    pages = "42--54",
    year = "2003"
}

@misc{SNO_water,
	title = {Low Energy Background in the NCD Phase of the Sudbury Neutrino Observatory},
% 	howpublished = {\url{https://sno.phy.queensu.ca/papers/OKEEFFE_THESIS.pdf}},
	note = {PhD Thesis, University of Oxford},
	author = {H. M. O'Keeffe}
}

@misc{Enomoto_Geo_Flux,
	title = {Neutrino Geophysics and Observation of Geo-Neutrinos at KamLAND},
% 	howpublished = {\url{https://sno.phy.queensu.ca/papers/OKEEFFE_THESIS.pdf}},
	note = {PhD Thesis, University of Tohoku},
	author = {E. Sanshiro}
}

@misc{KamLAND_preliminary,
	title = {KamLAND},
%  	howpublished = {\url{https://www.tfc.tohoku.ac.jp/wp-content/uploads/2016/10/04_HirokoWatanabe_TFC2016.pdf}},
	note = {International Workshop: Neutrino Research and Thermal Evolution of the Earth (2016)},
	author = {Hiroko Watanabe}
%	year = "2016"
}

@misc{Dye:2015bsw,
    author = "Dye, Steve and Barna, Andrew",
    howpublished = "{Global Antineutrino Modeling for a Web Application arxiv:1510.05633}",
    eprint = "1510.05633",
    archivePrefix = "arXiv",
    primaryClass = "physics.ins-det",
    month = "10",
    year = "2015"
}

@article{Muon_flux_Majorana,
title = {Muon flux measurements at the davis campus of the sanford underground research facility with the majorana demonstrator veto system},
journal = {Astroparticle Physics},
volume = {93},
pages = {70-75},
year = {2017},
issn = {0927-6505},
doi = {https://doi.org/10.1016/j.astropartphys.2017.01.013},
url = {https://www.sciencedirect.com/science/article/pii/S0927650517300038},
author = {N. Abgrall and E. Aguayo and F.T. Avignone and A.S. Barabash and F.E. Bertrand and A.W. Bradley and V. Brudanin and M. Busch and M. Buuck and D. Byram and A.S. Caldwell and Y-D. Chan and C.D. Christofferson and P.-H. Chu and C. Cuesta and J.A. Detwiler and C. Dunagan and Yu. Efremenko and H. Ejiri and S.R. Elliott and A. Galindo-Uribarri and T. Gilliss and G.K. Giovanetti and J. Goett and M.P. Green and J. Gruszko and I.S. Guinn and V.E. Guiseppe and R. Henning and E.W. Hoppe and S. Howard and M.A. Howe and B.R. Jasinski and K.J. Keeter and M.F. Kidd and S.I. Konovalov and R.T. Kouzes and B.D. LaFerriere and J. Leon and A.M. Lopez and J. MacMullin and R.D. Martin and R. Massarczyk and S.J. Meijer and S. Mertens and J.L. Orrell and C. O’Shaughnessy and N.R. Overman and A.W.P. Poon and D.C. Radford and J. Rager and K. Rielage and R.G.H. Robertson and E. Romero-Romero and M.C. Ronquest and C. Schmitt and B. Shanks and M. Shirchenko and N. Snyder and A.M. Suriano and D. Tedeschi and J.E. Trimble and R.L. Varner and S. Vasilyev and K. Vetter and K. Vorren and B.R. White and J.F. Wilkerson and C. Wiseman and W. Xu and E. Yakushev and C.-H. Yu and V. Yumatov and I. Zhitnikov}
}

@article{fast_neutrons_PhysRevD.73.053004,
  title = {Muon-induced background study for underground laboratories},
  author = {Mei, D.-M. and Hime, A.},
  journal = {Phys. Rev. D},
  volume = {73},
  issue = {5},
  pages = {053004},
  numpages = {18},
  year = {2006},
  publisher = {American Physical Society},
  doi = {10.1103/PhysRevD.73.053004},
  url = {https://link.aps.org/doi/10.1103/PhysRevD.73.053004}
}

@article{PhysRevLett.108.052505,
  title = {Analysis of $\ensuremath{\gamma}$-Ray Production in Neutral-Current Neutrino-Oxygen Interactions at Energies above 200 MeV},
  author = {Ankowski, Artur M. and Benhar, Omar and Mori, Takaaki and Yamaguchi, Ryuta and Sakuda, Makoto},
  journal = {Phys. Rev. Lett.},
  volume = {108},
  issue = {5},
  pages = {052505},
  numpages = {5},
  year = {2012},
  publisher = {American Physical Society},
  doi = {10.1103/PhysRevLett.108.052505},
  url = {https://link.aps.org/doi/10.1103/PhysRevLett.108.052505}
}

@article{Folomeshkin:1976vj,
    author = "Folomeshkin, V. N. and Gershtein, S. S and Khlopov, M. Yu. and Gmitro, M. and Eramzhian, R. A. and Tosunian, L. A.",
    title = "{Possible Tests of Neutral Current Models in the Neutrino Excitation of Light Atomic Nuclei}",
    doi = "10.1016/0375-9474(76)90668-0",
    journal = "Nucl. Phys. A",
    volume = "267",
    pages = "395--412",
    year = "1976"
}

@article{Birks:1951boa,
    author = "Birks, J. B.",
    title = "{Scintillations from Organic Crystals: Specific Fluorescence and Relative Response to Different Radiations}",
    doi = "10.1088/0370-1298/64/10/303",
    journal = "Proc. Phys. Soc. A",
    volume = "64",
    pages = "874--877",
    year = "1951"
}

@article{Bignell:2015oqa,
    author = "Bignell, Lindsey J. and others",
    title = "{Characterization and Modeling of a Water-based Liquid Scintillator}",
    eprint = "1508.07029",
    archivePrefix = "arXiv",
    primaryClass = "physics.ins-det",
    doi = "10.1088/1748-0221/10/12/P12009",
    journal = "JINST",
    volume = "10",
    number = "12",
    pages = "P12009",
    year = "2015"
}

@article{SNO:2020bdq,
    author = "Anderson, M. R. and others",
    collaboration = "SNO+",
    title = "{Measurement of neutron-proton capture in the SNO+ water phase}",
    eprint = "2002.10351",
    archivePrefix = "arXiv",
    primaryClass = "physics.ins-det",
    doi = "10.1103/PhysRevC.102.014002",
    journal = "Phys. Rev. C",
    volume = "102",
    number = "1",
    pages = "014002",
    year = "2020"
}

@article{Geo_KamLand,
  title = {Reactor on-off antineutrino measurement with KamLAND},
  author = {Gando, A. and Gando, Y. and Hanakago, H. and Ikeda, H. and Inoue, K. and Ishidoshiro, K. and Ishikawa, H. and Koga, M. and Matsuda, R. and Matsuda, S. and Mitsui, T. and Motoki, D. and Nakamura, K. and Obata, A. and Oki, A. and Oki, Y. and Otani, M. and Shimizu, I. and Shirai, J. and Suzuki, A. and Takemoto, Y. and Tamae, K. and Ueshima, K. and Watanabe, H. and Xu, B. D. and Yamada, S. and Yamauchi, Y. and Yoshida, H. and Kozlov, A. and Yoshida, S. and Piepke, A. and Banks, T. I. and Fujikawa, B. K. and Han, K. and O'Donnell, T. and Berger, B. E. and Learned, J. G. and Matsuno, S. and Sakai, M. and Efremenko, Y. and Karwowski, H. J. and Markoff, D. M. and Tornow, W. and Detwiler, J. A. and Enomoto, S. and Decowski, M. P.},
  collaboration = {KamLAND Collaboration},
  journal = {Phys. Rev. D},
  issue = {3},
  pages = {033001},
  numpages = {10},
  year = {2013},
  doi = {10.1103/PhysRevD.88.033001},
  url = {https://link.aps.org/doi/10.1103/PhysRevD.88.033001}
}

@misc{Hocker:2007ht,
    author = "Hocker, Andreas and others",
    title = "TMVA - Toolkit for Multivariate Data Analysis",
    url = "{https://doi.org/10.48550/arXiv.physics/0703039}",
    howpublished = "{arXiv:physics/0703039}",
    eprint = "physics/0703039",
    archivePrefix = "arXiv",
    reportNumber = "CERN-OPEN-2007-007",
    month = "3",
    year = "2007"
}

@article{Fiorentini:2007te,
    author = "Fiorentini, Gianni and Lissia, Marcello and Mantovani, Fabio",
    title = "{Geo-neutrinos and Earth's interior}",
    eprint = "0707.3203",
    archivePrefix = "arXiv",
    primaryClass = "physics.geo-ph",
    doi = "10.1016/j.physrep.2007.09.001",
    journal = "Phys. Rept.",
    volume = "453",
    pages = "117--172",
    year = "2007"
}

@article{Esteban:2020cvm,
    author = "Esteban, Ivan and Gonzalez-Garcia, M. C. and Maltoni, Michele and Schwetz, Thomas and Zhou, Albert",
    title = "{The fate of hints: updated global analysis of three-flavor neutrino oscillations}",
    eprint = "2007.14792",
    archivePrefix = "arXiv",
    primaryClass = "hep-ph",
    reportNumber = "IFT-UAM/CSIC-112, YITP-SB-2020-21",
    doi = "10.1007/JHEP09(2020)178",
    journal = "JHEP",
    volume = "09",
    pages = "178",
    year = "2020"
}

@article{PhysRevD.82.093006,
  title = {Combined analysis of KamLAND and Borexino neutrino signals from Th and U decays in the Earth's interior},
  author = {Fogli, G. L. and Lisi, E. and Palazzo, A. and Rotunno, A. M.},
  journal = {Phys. Rev. D},
  volume = {82},
  issue = {9},
  pages = {093006},
  numpages = {9},
  year = {2010},
  month = {Nov},
  publisher = {American Physical Society},
  doi = {10.1103/PhysRevD.82.093006},
  url = {https://link.aps.org/doi/10.1103/PhysRevD.82.093006}
}

@article{Geo_Borexino,
  author = {Agostini, M. and Altenm\"uller, K. and Appel, S. and Atroshchenko, V. and Bagdasarian, Z. and Basilico, D. and Bellini, G. and Benziger, J. and Bick, D. and Bonfini, G. and Bravo, D. and Caccianiga, B. and Calaprice, F. and Caminata, A. and Cappelli, L. and Cavalcante, P. and Cavanna, F. and Chepurnov, A. and Choi, K. and D'Angelo, D. and Davini, S. and Derbin, A. and Di Giacinto, A. and Di Marcello, V. and Ding, X. F. and Di Ludovico, A. and Di Noto, L. and Drachnev, I. and Fiorentini, G. and Formozov, A. and Franco, D. and Gabriele, F. and Galbiati, C. and Gschwender, M. and Ghiano, C. and Giammarchi, M. and Goretti, A. and Gromov, M. and Guffanti, D. and Hagner, C. and Hungerford, E. and Ianni, Aldo and Ianni, Andrea and Jany, A. and Jeschke, D. and Kumaran, S. and Kobychev, V. and Korga, G. and Lachenmaier, T. and Lasserre, T. and Laubenstein, M. and Litvinovich, E. and Lombardi, P. and Lomskaya, I. and Ludhova, L. and Lukyanchenko, G. and Lukyanchenko, L. and Machulin, I. and Mantovani, F. and Manuzio, G. and Marcocci, S. and Maricic, J. and Martyn, J. and Meroni, E. and Meyer, M. and Miramonti, L. and Misiaszek, M. and Montuschi, M. and Muratova, V. and Neumair, B. and Nieslony, M. and Oberauer, L. and Onillon, A. and Orekhov, V. and Ortica, F. and Pallavicini, M. and Papp, L. and Penek, \"O. and Pietrofaccia, L. and Pilipenko, N. and Pocar, A. and Raikov, G. and Ranalli, M. T. and Ranucci, G. and Razeto, A. and Re, A. and Redchuk, M. and Ricci, B. and Romani, A. and Rossi, N. and Rottenanger, S. and Sch\"onert, S. and Semenov, D. and Skorokhvatov, M. and Smirnov, O. and Sotnikov, A. and Strati, V. and Suvorov, Y. and Tartaglia, R. and Testera, G. and Thurn, J. and Unzhakov, E. and Vishneva, A. and Vivier, M. and Vogelaar, R. B. and von Feilitzsch, F. and Wojcik, M. and Wurm, M. and Zaimidoroga, O. and Zavatarelli, S. and Zuber, K. and Zuzel, G.},
  collaboration = {Borexino Collaboration},
  journal = {Phys. Rev. D},
  volume = {101},
  issue = {1},
  pages = {012009},
  numpages = {63},
  year = {2020},
  publisher = {American Physical Society},
  doi = {10.1103/PhysRevD.101.012009},
  url = {https://link.aps.org/doi/10.1103/PhysRevD.101.012009}
}

@article{PhysRevD.103.052004,
  title = {MeV-scale performance of water-based and pure liquid scintillator detectors},
  author = {Land, B. J. and Bagdasarian, Z. and Caravaca, J. and Smiley, M. and Yeh, M. and Orebi Gann, G. D.},
  journal = {Phys. Rev. D},
  volume = {103},
  issue = {5},
  pages = {052004},
  numpages = {20},
  year = {2021},
  publisher = {American Physical Society},
  doi = {10.1103/PhysRevD.103.052004},
  url = {https://link.aps.org/doi/10.1103/PhysRevD.103.052004}
}

@article{fn_thermalization,
author = { Michihira   FUJINO  and  Kenji   SUMITA },
title = {Measurements of Neutron Thermalization Time Constant of Light Water by Pulsed Neutron Method},
journal = {Journal of Nuclear Science and Technology},
volume = {7},
number = {6},
pages = {277-284},
year  = {1970},
publisher = {Taylor & Francis},
doi = {10.1080/18811248.1970.9734685},

URL = { 
        https://www.tandfonline.com/doi/abs/10.1080/18811248.1970.9734685
    
},
eprint = { 
        https://www.tandfonline.com/doi/pdf/10.1080/18811248.1970.9734685
    
}

}

@article{reines53,
author	= "Reines, F. and Cowan, C. L.",
title		= "{Detection of the free neutrino}",
journal	= {Phys. Rev.},
volume	= 92,
year		= 1953,
pages	= "830 - 831"}

@misc{IAEA,
     % author          = "{IAEA} Nuclear Data Section",
      note           = {{IAEA} Nuclear Data Section \href{https://www-nds.iaea.org/relnsd/NdsEnsdf/QueryForm.html}{IAEA database}}
}

@article{Jinping,
      author         = "Wan, Linyan and Hussain, Ghulam and Wang, Zhe and Chen,
                        Shaomin",
      title          = "{Geoneutrinos at Jinping: Flux prediction and oscillation
                        analysis}",
      journal        = "Phys. Rev.",
      volume         = "D95",
      year           = "2017",
      number         = "5",
      pages          = "053001",
      doi            = "10.1103/PhysRevD.95.053001",
      eprint         = "1612.00133",
      archivePrefix  = "arXiv",
      primaryClass   = "hep-ex",
      SLACcitation   = "%%CITATION = ARXIV:1612.00133;%%"
}

@article{SMIRNOV2019103712,
title = {Experimental aspects of geoneutrino detection: Status and perspectives},
journal = {Progress in Particle and Nuclear Physics},
volume = {109},
pages = {103712},
year = {2019},
issn = {0146-6410},
doi = {https://doi.org/10.1016/j.ppnp.2019.103712},
url = {https://www.sciencedirect.com/science/article/pii/S014664101930047X},
author = {O. Smirnov}
}

@article{gamma_n_cross_section,
  title = {Absolute cross section for the photodisintegration of deuterium},
  author = {Birenbaum, Y. and Kahane, S. and Moreh, R.},
  journal = {Phys. Rev. C},
  volume = {32},
  issue = {6},
  pages = {1825--1829},
  numpages = {0},
  year = {1985},
  month = {Dec},
  publisher = {American Physical Society},
  doi = {10.1103/PhysRevC.32.1825},
  url = {https://link.aps.org/doi/10.1103/PhysRevC.32.1825}
}

\end{document}